\documentclass[journal=jacsat,manuscript=communication,etalmode=truncate,keywords=true]{achemso}
\setkeys{acs}{articletitle=false,maxauthors=100,etalmode=truncate,keywords=true}
\usepackage{amsmath,amssymb,units,txfonts}
\usepackage{amsfonts}

\usepackage{graphicx}
\usepackage{MnSymbol}
\usepackage{helvet,wasysym}
\usepackage[usenames,dvipsnames]{color}
\usepackage{colortbl,cuted,multirow}
\newcommand{\epeak}{\varepsilon_{\textit{peak}}^{\textit{PDOS}}}
\newcommand{\eHOMOPDOS}{\varepsilon_{\textrm{HOMO}}^{\textit{PDOS}}}

\newcommand{\eF}{\varepsilon_{\textrm{F}}}
\newcommand{\Evac}{E_{\textit{vac}}}
\newcommand{\eTid}{\varepsilon_{\textrm{Ti}^{3+}}}
\newcommand{\eVBM}{\varepsilon_{\textrm{VBM}}}
\usepackage[utf8x]{inputenc}
\usepackage{xfrac}
\DeclareMathOperator{\Imag}{Im}
\DeclareMathOperator{\Real}{Re}

\definecolor{JPCCBlue}{RGB}{34,80,169}
\definecolor{ACSCatalysisBlue}{RGB}{79,115,194}
\definecolor{JCPCGreen}{RGB}{11,122,64}
\definecolor{StyleColor}{RGB}{79,115,194}
\definecolor{abstractcolor}{RGB}{236,244,248}
\makeatletter\newenvironment{abstractbox}{
   \begin{lrbox}{\@tempboxa}\begin{minipage}{0.988\textwidth}}{\end{minipage}\end{lrbox}
   \colorbox{abstractcolor}{\usebox{\@tempboxa}}
}\makeatother

\renewcommand*\authorfont{\flushleft}
\renewcommand*\affilfont{\mdseries\flushleft\textrm}
\renewcommand*\titlesize{\Large}
\renewcommand*\titlefont{\flushleft\sf\bfseries}
\def\refname{\sffamily\color{StyleColor}\large$\blacksquare$\normalsize\ 
REFERENCES}
\usepackage{titlesec}
\titleformat{\section}{\bfseries\sffamily\color{StyleColor}}{\thesection.~}{0pt}{}
\titleformat{\subsection}[runin]{\bfseries\sffamily\normalsize}{\indent\thesubsection.~}{0pt}{}[.]
\titlespacing{\subsection}{0pt}{0pt}{*1}
\titleformat{\subsubsection}{\bfseries\sffamily\normalsize}{\thethesubsection.~}{0pt}{}
\titlespacing{\subsubsection}{0pt}{0pt}{*0}

\newcommand{\newRev}[1]{#1}
\newcommand{\oldRev}[1]{}

\title{Comparing quasiparticle H$_{\text{2}}$O level alignment on anatase and rutile TiO$_{\text{2}}$}
\author{Huijuan Sun}
\affiliation[HFNLUSTC]{\newline\footnotemark[2]{\ } Hefei National Laboratory for Physical Sciences at the Microscale, University of Science and Technology of China, Hefei, Anhui 230026, China.} 
\alsoaffiliation[UPV/EHU]{\newline\footnotemark[3]{\ } Nano-Bio Spectroscopy Group and ETSF Scientific Development Center, Departamento de F{\'{\i}}sica de Materiales, Centro de F\'{\i}sica de Materiales CSIC-UPV/EHU-MPC and DIPC, Universidad del Pa{\'{\i}}s Vasco UPV/EHU, E-20018 San Sebasti\'{a}n, Spain}
\author{Duncan J.~Mowbray}
\email{duncan.mowbray@gmail.com}
\affiliation[UPV/EHU]{\newline\footnotemark[3]{\ } Nano-Bio Spectroscopy Group and ETSF Scientific Development Center, Departamento de F{\'{\i}}sica de Materiales, Centro de F\'{\i}sica de Materiales CSIC-UPV/EHU-MPC and DIPC, Universidad del Pa{\'{\i}}s Vasco UPV/EHU, E-20018 San Sebasti\'{a}n, Spain}
\author{Annapaola Migani}
\affiliation[ICN2]{\newline\footnotemark[5]{\ } ICN2 - Institut Catal\`{a} de Nanoci\`{e}ncia i Nanotecnologia, ICN2 Building, Campus UAB, E-08193 Bellaterra (Barcelona), Spain}
\alsoaffiliation[CSIC]{\newline\footnotemark[4]{\ } CSIC - Consejo Superior de Investigaciones Cient{\'{\i}}ficas, ICN2 Building, Campus UAB, E-08193 Bellaterra (Barcelona), Spain}
\author{Jin Zhao}
\affiliation[HFNLUSTC]{\newline\footnotemark[2]{\ } Hefei National Laboratory for Physical Sciences at the Microscale, University of Science and Technology of China, Hefei, Anhui 230026, China.} 
\alsoaffiliation[PHYUSTC]{\newline\footnotemark[6]{\ } Department of Physics and ICQD, University of Science and Technology of China, Hefei, Anhui 230026, China}
\alsoaffiliation[SICQI]{\newline$^\perp$Synergetic Innovation Center of Quantum Information \& Quantum Physics, University of Science and Technology of China, Hefei, Anhui 230026, China}
\author{Hrvoje Petek}
\affiliation[UP]{\newline$^\#$Department of Physics and Astronomy, University of Pittsburgh, Pittsburgh, Pennsylvania 15260, USA}
\author{Angel Rubio}
\affiliation[UPV/EHU]{\newline\footnotemark[3]{\ } Nano-Bio Spectroscopy Group and ETSF Scientific Development Center, Departamento de F{\'{\i}}sica de Materiales, Centro de F\'{\i}sica de Materiales CSIC-UPV/EHU-MPC and DIPC, Universidad del Pa{\'{\i}}s Vasco UPV/EHU, E-20018 San Sebasti\'{a}n, Spain}
\alsoaffiliation[MaxPlanck]{\newline\newRev{$^@$Max Planck Institute for the Structure and Dynamics of Matter, Luruper Chaussee 149, D-22761 Hamburg, Germany}}

\begin{document}

\maketitle

\begin{strip}
\vspace{-1.cm}

\noindent{\color{StyleColor}{\rule{\textwidth}{0.5pt}}}
\begin{abstractbox}
\begin{tabular*}{17cm}{b{8.26cm}r}
\noindent\textbf{\color{StyleColor}{ABSTRACT:}}
Knowledge of the molecular frontier levels' alignment in the ground state can be used to predict the photocatalytic activity of an interface.  The position of the adsorbate's highest occupied molecular orbital (HOMO) levels relative to the substrate's valence band maximum (VBM) in the interface describes the favorability of photogenerated hole transfer from the VBM to the adsorbed molecule. This is a key quantity for assessing and comparing H$_2$O photooxidation activities on two prototypical photocatalytic TiO$_2$ surfaces: anatase (A)-TiO$_2$(101) and rutile (R)-TiO$_2$(110).  Using the projected density of states (DOS) from state-of-the-art quasiparticle \oldRev{QP}\newRev{(QP)} $G_0W_0$ calculations, we assess the relative photocatalytic activity of intact and dissociated H$_2$O on coordinately unsaturated (Ti$_{\textit{cus}}$) sites of idealized stoichiomet- &
\includegraphics[height=4.76cm]{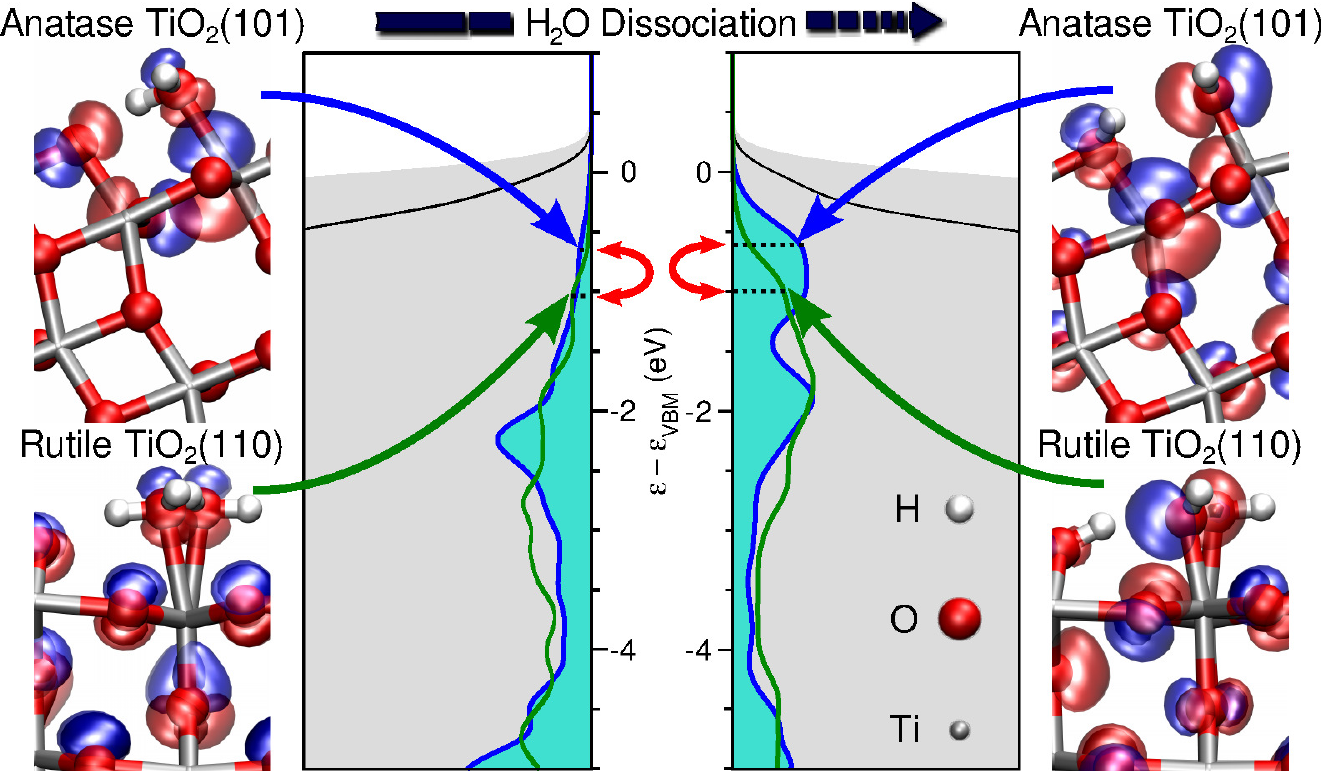}
\\
\multicolumn{2}{p{17cm}}{ric A-TiO$_2$(101)/R-TiO$_2$(110) and bridging O vacancies (O$_{\textit{br}}^{\textit{vac}}$) of defective A-TiO$_{2-x}$(101)/R-TiO$_{2-x}$(110)  surfaces ($x=\textrm{\sfrac{1}{4}},\textrm{\sfrac{1}{8}}$) for various coverages.  Such a many-body treatment is necessary to correctly describe the anisotropic screening of electron-electron interactions at a photocatalytic interface, and hence obtain accurate interfacial level alignments.    The more favorable ground state HOMO level alignment for A-TiO$_2$(101) may explain why the anatase polymorph shows higher photocatalytic activities than the rutile polymorph. Our results indicate that (1) hole trapping is more favored on A-TiO$_2$(101) than R-TiO$_2$(110) and (2) HO@Ti$_{\textit{cus}}$ is more photocatalytically active than intact H$_2$O@Ti$_{\textit{cus}}$.}\\
\multicolumn{2}{p{17cm}}{{\color{StyleColor}{\textbf{KEYWORDS:}}}\emph{ water splitting, $G_0W_0$ calculations, photocatalysis, photooxidation, hole trapping, titania}}
\end{tabular*}
\end{abstractbox}
\noindent{\color{StyleColor}{\rule{\textwidth}{0.5pt}}}
\end{strip}

\def\bigfirstletter#1#2{{\noindent
    \setbox0\hbox{{\color{StyleColor}{\Huge #1}}}\setbox1\hbox{#2}\setbox2\hbox{(}
    \count0=\ht0\advance\count0 by\dp0\count1\baselineskip
    \advance\count0 by-\ht1\advance\count0 by\ht2
    \dimen1=.5ex\advance\count0 by\dimen1\divide\count0 by\count1
    \advance\count0 by1\dimen0\wd0
    \advance\dimen0 by.25em\dimen1=\ht0\advance\dimen1 by-\ht1
    \global\hangindent\dimen0\global\hangafter-\count0
    \hskip-\dimen0\setbox0\hbox to\dimen0{\raise-\dimen1\box0\hss}
    \dp0=0in\ht0=0in\box0}#2}

\section{INTRODUCTION}
TiO$_2$ is widely used in photocatalysis, photoelectrocatalysis, and photovoltaics\cite{FujishimaReview,Diebold200353,TiO2PhotocatalysisChemRev2014,HendersonSurfSciRep,YatesChemRev,Setvin30082013}.  In particular, the H$_2$O--TiO$_2$ interface has been intensively studied both experimentally \cite{HendersonSurfSciRep2002} and theoretically\cite{SelloniWaterReview2010}. This is due to both the ubiquitous nature of the aqueous environment, and the technological importance of water splitting \cite{FujishimaNature,H2OandCO2reductionTiO2patents}. Because \oldRev{of the thermodynamic disadvantage to growing }large single-crystal samples of the anatase polymorph \newRev{are less stable than the rutile polymorph}\cite{PhysRevLett.81.2954,SunNatComm2013,Diebold2003CatalToday}, most surface science studies have focused on the rutile (110) surface of TiO$_2$.   However, \newRev{in the nanoparticle form} the anatase polymorph \oldRev{of TiO$_2$ yields}\newRev{is} more stable\oldRev{ nanoparticles},\cite{anataseNanopartical} \oldRev{which show}\newRev{and moreover it has a} higher photocatalytic \oldRev{activities}\newRev{activity} \cite{GratzelJACS1996}. 

A proper assessment of an interface's photocatalytic activity requires an accurate description of its frontier levels' alignment.  This is because interfacial electron transfer is controlled by the alignment of the highest occupied and lowest unoccupied molecular orbitals (HOMO/LUMO) relative to the valence band maximum (VBM) and conduction band minimum (CBM) \cite{HendersonSurfSciRep,YatesChemRev}.  In particular, H$_2$O photooxidation, i.e., the oxygen evolution reaction (OER), is initiated by the transfer of the photogenerated hole from the substrate's VBM to the HOMO\cite{LiJACSPOE,SelloniPCETJACS2013,WaterDissociationJACS2012Jin}.  

\newRev{Based on the HOMO's position relative to the VBM prior to irradiation, i.e., vertical alignment, one may establish trends in photocatalytic activity among a group of systems\cite{FriendChemRec2014,ZungerPCCP2014}. Even in cases where the HOMO initially lies below the VBM\cite{MiganiH2O}, after light absorption and nuclear relaxation, these levels may reorder, with the hole localized on the molecule\cite{SprikSchematic}.   Essentially, the closer to the VBM and more localized on the molecule the HOMO is initially, the greater the molecule's propensity for trapping the hole.}
  For \oldRev{this}\newRev{these} reason\newRev{s}, the \oldRev{occupied interfacial levels' }alignment \newRev{of the H$_2$O occupied levels} prior to irradiation is most relevant for understanding the \oldRev{process of water splitting}\newRev{OER}.   

   Recently, we applied many-body quasiparticle (QP) $GW$ techniques\cite{GW,AngelGWReview} to determine the H$_2$O occupied levels' alignment on rutile (R)-TiO$_2$(110)\cite{MiganiH2O}.  We found that an accurate description of the interfacial anisotropic screening via QP $GW$ is essential to accurately describe the interfacial level alignment\cite{OurJACS,MiganiLong,MiganiInvited,MiganiH2O}.  Specifically, the occupied QP density of states (DOS) projected onto the molecule is an effective means for interpreting  difference spectra\oldRev{ obtained}\newRev{, i.e., the difference between spectra with a chemisorbed molecular overlayer and a clean TiO$_2$ surface,} from ultraviolet photoemission spectroscopy (UPS)\cite{MiganiH2O}.  Such theoretical approaches are necessary to disentangle highly hybridized adsorbate levels from those of the substrate, such as those of the H$_2$O--TiO$_2$ interface
\cite{MiganiH2O}.  

Here, we investigate the H$_2$O occupied levels' alignment on the anatase (A)-TiO$_2$(101) surface, as it is the most common surface in nanostructured TiO$_2$ \cite{Diebold200353,SelloniChemRev2014,SelloniPRB2001}.  
In the absence of UPS measurements for H$_2$O on A-TiO$_2$(101), we compare the results to the \oldRev{QP }$G_0W_0$ PDOS of H$_2$O on R-TiO$_2$(110) \cite{RutileVSAnatasePCCP2013}, which is consistent with UPS difference spectra\cite{ThorntonH2ODissTiO2110,DeSegovia,Krischok}.

\begin{figure}
\noindent{\color{StyleColor}{\rule{\columnwidth}{1pt}}}
\includegraphics[width=\columnwidth]{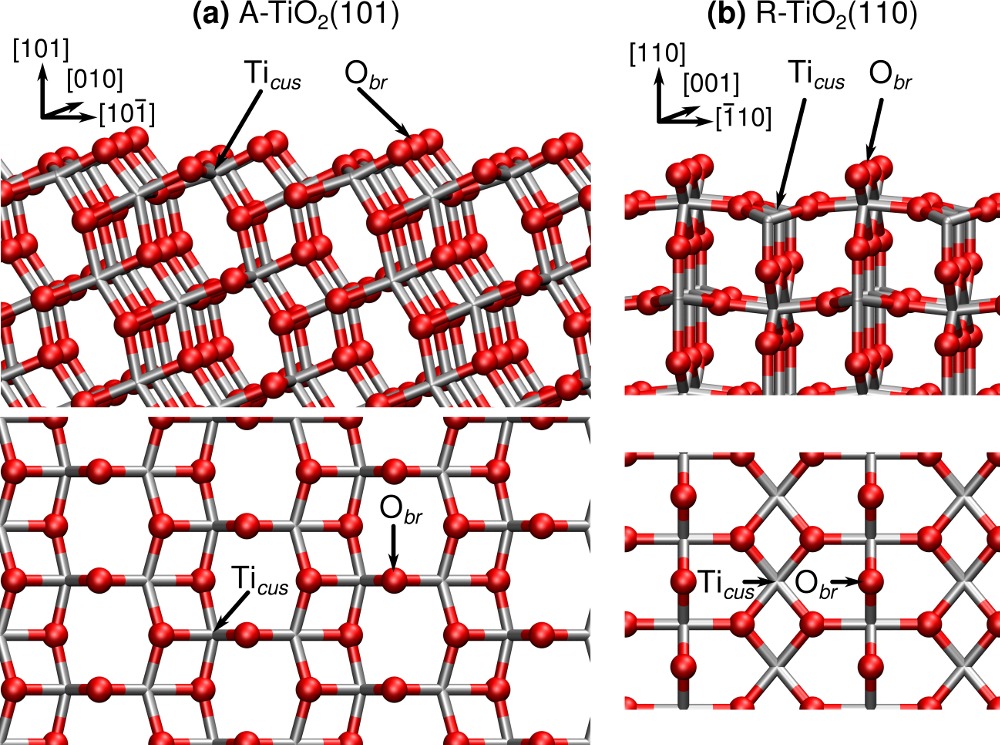}
\caption{Schematics of clean stoichiometric \textbf{(a)} A-TiO$_2$(101) and \textbf{(b)} R-TiO$_2$(110) surfaces.  Ti and O atoms are depicted in silver and red, respectively. Coordinately unsaturated Ti sites (Ti${_\textit{cus}}$) and bridging O atoms (O$_\textit{br}$) are labelled.}\label{fgr:CleanSurfaceSchematic}
\noindent{\color{StyleColor}{\rule{\columnwidth}{1pt}}}
\end{figure}

In particular, we perform \oldRev{QP single-shot }$G_0W_0$\cite{GW,AngelGWReview,KresseG0W0} and partially self-consistent\cite{KressescGW} \newRev{(sc)}\oldRev{sc}QP$GW1$\cite{OurJACS,MiganiLong} calculations based on Kohn-Sham (KS) levels from density functional theory (DFT) using a \newRev{local density approximation (LDA)\cite{LDA},} generalized gradient approximation (PBE)\cite{PBE}\newRev{,} or a range-separated hybrid (HSE)\cite{HSE,HSE06} exchange correlation (xc)-functionals.  From \oldRev{this}\newRev{these calculations} we obtain the total and projected QP DOS for a variety of coverages \oldRev{(}\newRev{[}\sfrac{1}{4} to 1\sfrac{1}{2} monolayer (ML)\oldRev{)}\newRev{]} of intact and dissociated H$_2$O adsorbed on coordinately unsaturated Ti sites (H$_2$O@Ti$_{\textit{cus}}$) of stoichiometric A-TiO$_2$(101) and bridging O vacancies (H$_2$O@O$_{\textit{br}}^{\textit{vac}}$) of defective A-TiO$_{2-\text{\sfrac{1}{4}}}$(101) and A-TiO$_{2-\text{\sfrac{1}{8}}}$(101) surfaces with \sfrac{1}{2}ML and \sfrac{1}{4}ML O$_{\textit{br}}^{\textit{vac}}$. The Ti$_{\textit{cus}}$ and O$_{\textit{br}}$ sites of A-TiO$_2$(101) and R-TiO$_2$(110) are shown schematically in Figure~\ref{fgr:CleanSurfaceSchematic}.

On the one hand, by considering the absolute interfacial level alignment\newRev{, i.e.,} relative to the vacuum level $\Evac$, one obtains the interface's ionization potential $\textit{IP} = -\eVBM + \Evac$.  This is the quantity that can be compared with red-ox potentials \cite{SprikPCCPREview2012,SprikH2OAlignment}.   Moreover, the absolute level alignment allows a direct comparison between alignments across different substrates\cite{ZungerPCCP2014}, such as A-TiO$_2$(101) and R-TiO$_2$(110).  Finally, from the absolute level alignment, one can determine whether changes in the H$_2$O level alignment across different substrates are attributable to shifts of the substrate or molecular levels.  

On the other hand, by considering the level alignment relative to the VBM of the substrate $\eVBM$, one may directly compare the favorability of photogenerated hole transfer from the substrate's VBM to the molecule's HOMO.  In combination with the $\textit{IP}$ this allows a robust comparison of photocatalytic activity across substrates. Moreover, by referencing the spectra to the VBM, one may directly compare the shape and dispersion of the valence band edge at the VBM. Finally, the VBM is the most reliable KS energy reference\newRev{, from a theoretical perspective\cite{MiganiLong}}.

For these reasons, we shall make use of both VBM and $E_{\textit{vac}}$ energy references as appropriate.  In particular, we provide the absolute level alignment when comparing to HSE DFT and HSE $G_0W_0$ DOS.  This is because the VBM and CBM from HSE DFT for clean\cite{MiganiLong} and 1ML H$_2$O@Ti$_{\textit{cus}}$\cite{SprikPCCPREview2012,SprikH2OAlignment,MiganiH2O} on R-TiO$_2$(110) are consistent with measurements for clean quasi-stoichiometric \cite{MiganiLong,TiO2WorkFunction30,Onishi198833,C0CP02835E,Petek2PPH2O} and liquid H$_2$O covered R-TiO$_2$(110)\cite{SprikH2OAlignment}, respectively.

We begin by providing a detailed description of the techniques\oldRev{ and}\newRev{,} parameters\newRev{, and terminology} employed throughout in Section \ref{Sect:Methodology}.  \newRev{In Section~\ref{Sect:Bulk101SurfaceATiO2} we compare our results to calculated and measured electronic band gaps $E_g$, optical band gaps $\hbar\omega_g$, and macroscopic dielectric constants $\varepsilon_\infty$ of bulk A-TiO$_2$ and R-TiO$_2$  available in the literature.}  To \newRev{further }test the robustness of our approach, and the parameters employed, we compare the dielectric function obtained by solving the Bethe-Salpeter equation (BSE) based on $G_0W_0$ eigenvalues with that obtained from reflection spectra measurements for bulk A-TiO$_2$\oldRev{ in Section \ref{Sect:Bulk101SurfaceATiO2}}.  We also compare the $\textit{IP}$  for clean idealized stoichiometric A-TiO$_2$(101) and R-TiO$_2$(110).  In Section \ref{Sect:Intact} we provide a detailed comparison of the H$_2$O PDOS for intact and \sfrac{1}{2} dissociated H$_2$O@Ti$_{\textit{cus}}$ on A-TiO$_2$(101) and R-TiO$_2$(110) relative to $\Evac$ and $\eVBM$ with PBE DFT, HSE DFT, PBE scQP$GW1$, PBE $G_0W_0$ and HSE $G_0W_0$.  In Section \ref{Sect:Dissociated} we provide a similar detailed comparison for dissociatively adsorbed H$_2$O@O$_{\textit{br}}^{\textit{vac}}$ on A-TiO$_{2-\textrm{\sfrac{1}{4}}}$(101) and R-TiO$_{2-\textrm{\sfrac{1}{4}}}$(110) relative to $\Evac$.  Finally, in Section \ref{Sect:Dependence} we extend the comparison of A-TiO$_2$(101) and R-TiO$_2$(110) to their adsorption energies and level alignments relative to $\eVBM$ with H$_2$O coverage (\sfrac{1}{4} to 1\sfrac{1}{2}ML), H$_2$O dissociation (intact to fully dissociated), and surface composition (O$_{\textit{br}}^{\textit{vac}}$ coverage).  This is followed by concluding remarks.

\section{METHODOLOGY}\label{Sect:Methodology}

\subsection{\newRev{Theoretical Methods}}
\newRev{DFT based on standard xc-functionals, e.g., LDA and PBE, tends to significantly underestimate the electronic band gaps of semiconducting metal oxides, such as TiO$_2$\cite{TiO2GWVasp}.  This is in part due to their underestimation of the screening of the electron-electron interaction.}  

\newRev{DFT based on hybrid xc-functionals, e.g., HSE,  partially remedies this by replacing a fraction of the exchange term with Hartree-Fock exact-exchange.  The fraction of Hartree-Fock exact-exchange included, $\alpha$, acts as an effective constant screening of the Hartree-Fock electron-electron interaction term, i.e., an inverse dielectric constant for the system $\alpha \sim \varepsilon_\infty^{-1}$.\cite{Marques}    In particular, we use the HSE06 variant, with a \oldRev{screening}\newRev{range separation} parameter of $\mu = 0.2$~\AA$^{-1}$, of the HSE hybrid xc-functional, which includes 25\% exact-exchange ($\alpha = 0.25$)\cite{HSE}.  For systems where the screening is rather homogeneous, and $\varepsilon_\infty \sim 4$,  this leads to a better description of the electronic band gap\cite{Marques}, e.g., of bulk TiO$_2$\cite{TiO2GWVasp}.}  

\newRev{However, since HSE applies the same screening to all the levels regardless of their nature, it fails to describe the anisotropic screening felt by molecular levels at an interface.  As a result, localized occupied molecular levels are underbound by  HSE\cite{OurJACS,MiganiLong,MiganiH2O}. This can lead to significant errors in HSE's description of molecular/semiconductor \oldRev{the }interfacial level alignment\cite{OurJACS,MiganiLong,MiganiH2O}. Instead, QP techniques, e.g., $G_0W_0$ and scQP$GW1$, where the spacial dependence of the screening is explicitly included, provide a better description of the interfacial level alignment.\cite{OurJACS,MiganiLong,MiganiInvited,MiganiH2O}}

\newRev{In the \oldRev{QP }$G_0W_0$ approach, the contribution to the KS eigenvalues from the xc-potential $V_{xc}$ is replaced by the self   energy $\Sigma = i G W$, where $G$ is the Green's function and $W$ is the spatially dependent screening \cite{GW} obtained from the KS wavefunctions \cite{AngelGWReview}.  The dielectric function is obtained from linear response time-dependent (TD) DFT within the random phase approximation (RPA), including local field effects \cite{KresseG0W0}.  From $G_0W_0$ one obtains first-order QP corrections to the KS eigenvalues, but retains the KS wavefunctions.  Generally, these QP corrections to the occupied levels are linearly dependent on the fraction of the KS wavefunction's density in the molecular layer \cite{OurJACS,MiganiLong}.  This means the screening of these levels $W$ is quite anisotropic.  For this reason, QP $GW$ methods are necessary to accurately describe the interfacial level alignment.}

\oldRev{Our \oldRev{QP }$G_0W_0$ calculations\cite{GW,AngelGWReview,KresseG0W0} have been
performed using \textsc{vasp} within the projector augmented wave (PAW) scheme
\cite{kresse1999}.   The $G_0W_0$ calculations are based on \oldRev{Kohn-Sham}\newRev{KS}
wavefunctions and eigenenergies obtained from DFT using either PBE \cite{PBE} or HSE \cite{HSE} xc-functionals \cite{kresse1996b}.}

\newRev{Moreover, to include dependencies on the QP wavefunctions, and possibly obtain an improved absolute level alignment for the interface, one can also employ self-consistent QP techniques, such as scQP$GW1$\cite{OurJACS,MiganiLong,MiganiH2O}.}
\oldRev{We have also}\newRev{Here, we have} employed a single-shot \newRev{sc}QP\oldRev{ }$GW$\newRev{1} approach\cite{KressescGW,OurJACS}, where 100\% of the DFT xc-potential is replaced by the QP self energy in a single self-consistent \newRev{sc}QP\oldRev{ }$GW$ cycle. \newRev{We employ this procedure to make practical scQP$GW$ calculations for large interfaces, such as H$_2$O--A-TiO$_2$(101).}  In so doing, one obtains eigenvalues comparable to those from \oldRev{QP }$G_0W_0$\newRev{, along with the QP wavefunctions}.   This differs from the scQP$GW1$ approach \newRev{as} previously applied to the H$_2$O--R-TiO$_2$(110) interface\cite{MiganiH2O}, where 25\%, 25\%, and 50\%, of the QP self energies were ``mixed'' with the DFT xc-potential over three \oldRev{self-consistent QP }\newRev{scQP}$GW$ cycles \cite{KressescGW}, respectively. 

\subsection{\newRev{Computational Details}}
\newRev{Our \oldRev{QP }$G_0W_0$ calculations\cite{GW,AngelGWReview,KresseG0W0} have been
performed using \textsc{vasp} within the projector augmented wave (PAW) scheme
\cite{kresse1999}.   The $G_0W_0$ calculations are based on \oldRev{Kohn-Sham}\newRev{KS}
wavefunctions and eigenenergies obtained from DFT
using either LDA\cite{LDA}, PBE\cite{PBE}, or HSE \cite{HSE} xc-functionals \cite{kresse1996b}.}

The geometries have been fully relaxed using \newRev{LDA\cite{LDA} or} PBE \cite{PBE}, with all forces $\lesssim$ 0.02 eV/\AA. The HSE\cite{HSE} calculations have been performed for the relaxed geometries obtained with PBE. We have employed a plane-wave energy cutoff of 445 eV, an electronic temperature of $k_B T\approx0.1$ eV with all energies extrapolated to $T\rightarrow 0$ K, and a PAW pseudopotential for Ti which includes the 3$s^2$ and 3$p^6$ semi-core levels.  All calculations have been performed spin unpolarized\cite{MiganiH2O}.

For the clean stoichiometric A-TiO$_2$(101) surface we have used a five-layer slab, an orthorhombic  $1\times1$ unit cell of $10.23 \times 3.78\times 40$ \AA$^3$, a $\Gamma$-centered $4\times 8\times1$ \textbf{k}-point mesh, and approximately 9\sfrac{1}{3} unoccupied bands per atom.  For the clean defective A-TiO$_{2-\text{\sfrac{1}{4}}}$(101) surface we have used a monoclinic $1\times2$ unit cell of $10.23 \times 7.56\times 40$ \AA$^3$ and a $\Gamma$-centered $4\times 4\times1$ \textbf{k}-point mesh.  For the clean defective A-TiO$_{2-\text{\sfrac{1}{8}}}$(101) surface we have used a $1\times4$ unit cell of $10.23 \times 15.13\times 40$ \AA$^3$ and a $\Gamma$-centered $4\times 2\times1$ \textbf{k}-point mesh.  For the H$_2$O covered surfaces, we have employed a five-layer slab with  adsorbates on both sides, an orthorhombic $1\times 1$ unit cell of $10.23  \times 3.78 \times 47$~\AA$^3$, a $\Gamma$ centered $4\times8\times1$ \textbf{k}-point mesh, and approximately 9\sfrac{1}{6} unoccupied bands per atom, i.e., including all levels up to 30~eV above the VBM, an energy cutoff of 80 eV for the number of \textbf{G}-vectors, and a sampling of 80 frequency points for the \newRev{RPA }dielectric function\oldRev{ within the random phase approximation (RPA)}.   The $G_0W_0$ parameters are consistent with those previously  used  for describing bulk R-TiO$_2$, R-TiO$_2$(110) clean surface and interfaces\cite{OurJACS,MiganiLong}.  Although our \oldRev{QP }$G_0W_0$ calculations do not include electron-phonon\cite{RinkeSemiSciTech2011} and lattice polarization\cite{MarquesLatticePolarization} contributions, these parameters have been shown to provide accurate descriptions of bulk optical absorption spectra, and both clean surface and interfacial level alignment\cite{OurJACS,MiganiLong}.

It has previously been shown\cite{AmilcareTiO2GW,HybertsenTiO2GW,TiO2GWVasp} that the experimental optical spectra for bulk A-TiO$_{\text{2}}$ may be obtained via BSE \cite{KresseBSE} based on $G_0W_0$ eigenvalues. In our BSE calculations, we include the electrostatic electron-hole interaction using the effective nonlocal frequency independent exchange correlation $f_{xc}({\textbf{r}},{\textbf{r}}',\omega=0)$ kernel suggested in ref.~\citenum{Reiningfxc}\nocite{Reiningfxc}.   
For \oldRev{the BSE calculations of }bulk A-TiO$_2$, we have used a tetragonal conventional 12 atom supercell with experimental lattice parameters $a=b=3.78$~\AA\ and $c=9.5$~\AA\cite{TiO2LatticeParameters}, and a dense $\Gamma$-centered $10\times10\times4$ \textbf{k}-point mesh\newRev{. For bulk R-TiO$_2$, we have used a tetragonal 6 atom primitive cell with experimental lattice parameters $a=b=4.5941$~\AA\ and $c=2.958$~\AA\cite{TiO2LatticeParameters}, a $\Gamma$-centered $6\times6\times10$ \textbf{k}-point mesh with PBE and HSE and a denser $\Gamma$-centered $8\times8\times12$ \textbf{k}-point mesh with LDA}\oldRev{ with 480 sampling points for the RPA dielectric function}.  \oldRev{We}\newRev{For both A-TiO$_2$ and R-TiO$_2$, we} have included $n_{\textrm{unocc}} = 12$ unoccupied bands per atom\oldRev{ for the $G_0W_0$ and }\newRev{. For the BSE calculations of bulk A-TiO$_2$, we have used 480 sampling points for the RPA dielectric function, and included} all the transitions between the 16 highest energy occupied bands and the 12 lowest energy unoccupied bands\oldRev{ in the BSE calculation}.\cite{KresseBSE}

\subsection{\newRev{Terminology}}
\oldRev{In this study}\newRev{To compare the relative stabilities of the H$_2$O covered anatase and rutile polymorphs}, we have performed single-point RPBE\cite{RPBE} based DFT calculations using the PBE relaxed structure for the H$_2$O adsorption energies $E_{\textit{ads}}$ on \oldRev{the }stoichiometric \newRev{A-TiO$_2$(101)} and defective \newRev{A-TiO$_{2-x}$(101)} surfaces.  The RPBE xc-functional was especially developed for the prediction of adsorption properties on metal surfaces \cite{RPBE}.   Furthermore, RPBE has been shown to provide accurate formation energies for metal dioxides\cite{MartinezMO2} and perovskites \cite{CalleVallejoPerovskites}. 

The H$_2$O adsorption energy on the Ti$_{\textit{cus}}$ site of a stoichiometric \newRev{A-}TiO$_2$(101) surface is given by
\begin{equation}
E_{\textit{ads}} \approx \frac{E[n\textrm{H}_2\textrm{O}+\textrm{\newRev{A-}TiO}_2\textrm{(101)}] - E[\textrm{\newRev{A-}TiO}_2\textrm{(101)}]}{n}- E[\textrm{H}_2\textrm{O}],
\end{equation}
where $n$ is the number of adsorbed H$_2$O functional units in the supercell,
and $E[n\textrm{H}_2\textrm{O}+\textrm{\newRev{A-}TiO}_2\textrm{(101)}]$,
$E[\textrm{\newRev{A-}TiO}_2\textrm{(101)}]$, and $E[\textrm{H}_2\textrm{O}]$ are the total
energies of the covered and clean stoichiometric surfaces and gas phase water
molecule, respectively.  Similarly, the H$_2$O adsorption energy on the
O$_{\textit{br}}^{\textit{vac}}$ site of a defective \newRev{A-}TiO$_{2-x}$(101) surface is given by
\begin{equation}
E_{\textit{ads}} \approx \frac{E[n\textrm{H}_2\textrm{O}+\textrm{\newRev{A-}TiO}_{2-x}\textrm{(101)}] - E[\textrm{\newRev{A-}TiO}_{2-x}\textrm{(101)}]}{n} - E[\textrm{H}_2\textrm{O}] ,
\end{equation}
where $E[n\textrm{H}_2\textrm{O}+\textrm{\newRev{A-}TiO}_{2-x}\textrm{(101)}]$ and $E[\textrm{\newRev{A-}TiO}_{2-x}\textrm{(101)}]$ are the total energies of the covered and clean defective surfaces, respectively.

\newRev{To provide a quantitative comparison between the DOS for the H$_2$O--A-TiO$_2$ and H$_2$O--R-TiO$_2$ interfaces, we employ the interfaces' $\textit{IP}$s. These are obtained from the difference in energy between the vacuum level $\Evac$ and the VBM $\eVBM$, $\textit{IP} = -\eVBM + \Evac$, where $\Evac$ is the maximum surface averaged electrostatic potential in the vacuum region between slabs.}

\newRev{Similarly, to provide a quantitative comparison between the PDOS for the H$_2$O--A-TiO$_2$ and H$_2$O--R-TiO$_2$ interfaces, we employ both the highest H$_2$O PDOS peak $\epeak$ and the average energy of the highest energy electron, or HOMO, of the PDOS, $\eHOMOPDOS$. To obtain $\epeak$ from the PDOS, we fit three Gaussians to the first few peaks below the VBM.  In this way we may disentangle the highest energy peak when it forms a shoulder within the upper edge of the PDOS.}

\newRev{However, to assess trends in the comparative photocatalytic activity of the H$_2$O--A-TiO$_2$ and H$_2$O--R-TiO$_2$ interfaces, one should consider not only a peak's energy, but also differences in its' intensity, i.e., localization on H$_2$O. Both quantities are incorporated within the single descriptor $\eHOMOPDOS$.  
We define $\eHOMOPDOS$ as the first moment of the PDOS, $\rho^{\textit{PDOS}}(\varepsilon)$ over the interval encompassing the highest energy electron.  More precisely,}
\begin{equation}
\newRev{\eHOMOPDOS \equiv \int_{E_1}^{\eVBM + \Delta}  \varepsilon \rho^{\textit{PDOS}}\!\!(\varepsilon)d \varepsilon,}
\end{equation}
\newRev{where $\eVBM$ is the VBM energy, $\Delta \sim 1$~eV ensures the tail of the VBM is included within the integral, and $E_1$ is the lower bound of the energy range encompassing the highest energy electron of the PDOS, i.e.,}
\begin{equation}
\newRev{\int_{E_1}^{\eVBM + \Delta}  \rho^{\textit{PDOS}}\!\!(\varepsilon)d \varepsilon \equiv 1.}
\end{equation}

\section{RESULTS AND DISCUSSION}

\subsection{Bulk and (101) Surface of Anatase TiO$_{\text{2}}$}\label{Sect:Bulk101SurfaceATiO2}

To test the reliability of the parameters we have employed to
calculate the \oldRev{QP }$G_0W_0$ levels of A-TiO$_2$, we first consider the optical
response of bulk anatase. Previous DFT band structure calculations\cite{AmilcareTiO2GW,HybertsenTiO2GW,TiO2GWVasp} found A-TiO$_2$ has an indirect electronic band gap between the \oldRev{CBM at $\Gamma$ and the }VBM along the $\Sigma$ path at $0.88 \Gamma\rightarrow\textrm{M}$\cite{TiO2GWVasp}\newRev{, i.e., $\Sigma$, and the CBM at $\Gamma$}. Our
PBE $G_0W_0$ calculation yields an indirect electronic band gap for A-TiO$_2$ of 3.8\oldRev{5}\newRev{6} eV, from a VBM at $0.8 \Gamma\rightarrow\textrm{M}$.
This is comparable with the \oldRev{QP }$G_0W_0$ indirect band gaps reported in the literature\newRev{, as shown in Table~\ref{Eg:tbl}}\oldRev{\cite{AmilcareTiO2GW,HybertsenTiO2GW,TiO2GWVasp}}.

\begin{table}
\vspace{-9mm}
\noindent{\color{StyleColor}{\rule{\columnwidth}{1.0pt}}}
\vspace{4.5mm}
\caption{\newRev{\textrm{\bf Direct and Indirect Band Gaps $\boldsymbol{E_{g}}$ and Optical Gaps  $\boldsymbol{\hslash\omega_g}$ in eV of A-TiO$_{\text{2}}$ and R-TiO$_{\text{2}}$.}}
}\label{Eg:tbl}
\begin{tabular}{ccllll}
\multicolumn{6}{>{\columncolor[gray]{0.9}}c}{ }\\[-3mm]
\multicolumn{1}{>{\columncolor[gray]{0.9}}c}{method} &
\multicolumn{1}{>{\columncolor[gray]{0.9}}c}{xc-functional} &
\multicolumn{2}{>{\columncolor[gray]{0.9}}c}{A-TiO$_2$} &
\multicolumn{2}{>{\columncolor[gray]{0.9}}c}{R-TiO$_2$} \\
& &
\multicolumn{4}{c}{electronic band gap} \\
\multicolumn{2}{c}{} 
 &\multicolumn{1}{l}
{$\Gamma\rightarrow\Gamma$} 
&\multicolumn{1}{l}
{$\Sigma\rightarrow \Gamma$} 
&\multicolumn{1}{l}
{$\Gamma\rightarrow\Gamma$} 
&\multicolumn{1}{l}
{$\Gamma\rightarrow R$} \\\cline{3-6}\\[-3mm]
\multirow{2}{*}{DFT} & \multirow{2}{*}{HSE} &  3.72 & 3.63 & 3.40 & 3.40 \\
&   & & 3.60$^a$ & 3.39$^a$ & 3.39$^a$ \\
\multirow{4}{*}{$G_0W_0$} & \multirow{2}{*}{LDA} & 3.93 & 3.86 & 3.33 & 3.26\\
& & 4.14$^b$ & 3.56$^b$ & 3.38$^b$ & 3.34$^b$\\
& \multirow{2}{*}{PBE} & & 3.73$^a$ & 3.46$^a$\\
& & 4.29$^c$ & 3.83$^c$ & 3.59$^c$ \\
& PBE+$\Delta$ & & 3.57$^d$ & 3.30$^d$ & 3.23$^d$ \\
\multicolumn{2}{c}{PES/IPES} & & & \multicolumn{2}{c}{$3.3\pm 0.5^e$}\\
\multicolumn{2}{c}{$(\alpha_{\textrm{KM}}\cdot \hslash\omega)^2$} & \multicolumn{2}{c}{3.53$^f$} & \multicolumn{2}{c}{3.37$^f$}\\
\multicolumn{2}{c}{}& 
\multicolumn{4}{c}{optical gap}\\\cline{3-6}\\[-3mm]
\multirow{2}{*}{BSE} & LDA & \multicolumn{2}{c}{3.73$^{\ }$} & \multicolumn{2}{c}{3.15$^{\ }$}\\
 & PBE & \multicolumn{2}{c}{3.57$^a$} & \multicolumn{2}{c}{3.28$^a$}\\
\multicolumn{2}{c}{Transmission} &  \multicolumn{2}{c}{3.42$^g$}\\
\multicolumn{2}{c}{Absorption}
& & & \multicolumn{2}{c}{3.03$^h$}\\
\multicolumn{2}{c}{Reflectance} &  \multicolumn{2}{c}{3.21$^i$} & \multicolumn{2}{c}{3.00$^i$
}\\[1mm]
\multicolumn{6}{p{0.95\columnwidth}}{\tiny$^a$Ref.~\citenum{TiO2GWVasp}\nocite{TiO2GWVasp}. 
$^b$Ref.~\citenum{HybertsenTiO2GW}\nocite{HybertsenTiO2GW}. $^c$Ref.~\citenum{AmilcareTiO2GW}\nocite{AmilcareTiO2GW}. $^d$Ref.~\citenum{TiO2OpticalPropertiesBSEDFT+U}\nocite{TiO2OpticalPropertiesBSEDFT+U}. $^e$Photoemission and Bremsstrahlung isochromat spectroscopy from ref.~\citenum{TiO2BandGap}\nocite{TiO2BandGap}. $^f$Estimate assuming a nearly direct band gap based on Kubelka–Munk adsorption coefficients $\alpha_{\textrm{KM}}$ from reflectance measurements of phase-pure nanoparticles in ref.~\citenum{TiO2NPCoronadoNanotech2008}\nocite{TiO2NPCoronadoNanotech2008}. $^g$Ref.~\citenum{UrbachTailTiO2}\nocite{UrbachTailTiO2}.
$^h$Refs.~\citenum{RutileOpticalGap} and \citenum{PhysRevB.18.5606}\nocite{RutileOpticalGap,PhysRevB.18.5606}.  $^i$For pure-phase nanoparticles from ref.~\citenum{TiO2NPCoronadoNanotech2008}.
}
\end{tabular}
{\color{StyleColor}{\rule{\columnwidth}{1.0pt}}}
\end{table}

\begin{figure}
\includegraphics[width=\columnwidth]{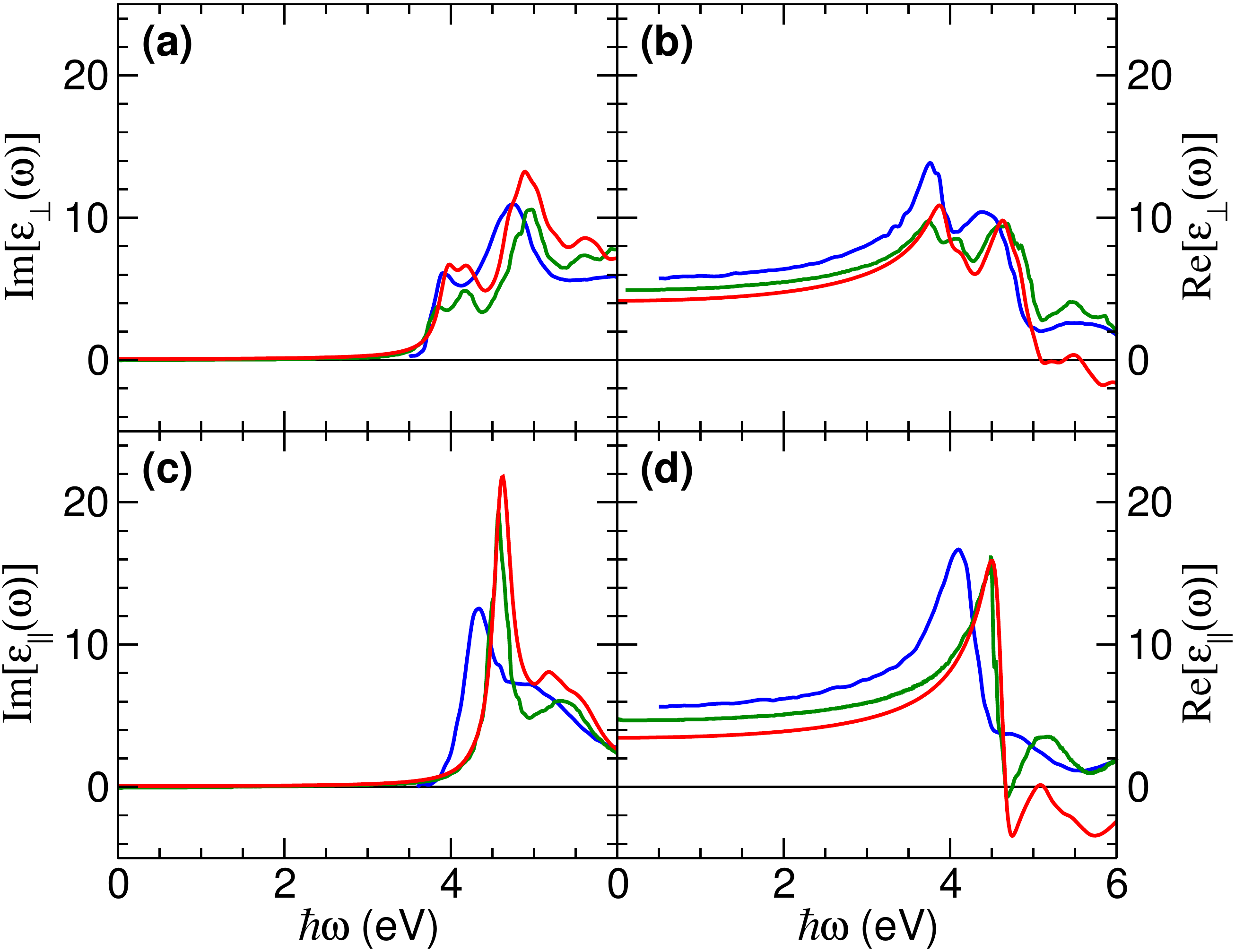}
\caption{\textbf{(a,c)} Imaginary and \textbf{(b,d)} real parts of the dielectric function of bulk A-TiO$_2$ for polarization perpendicular \textbf{(a,b)} and parallel \textbf{(c,d)} to the A-TiO$_2$ tetragonal $c$-axis, $\Imag [ \varepsilon_\perp(\omega)]$, $\Real [ \varepsilon_\perp(\omega)]$, $\Imag [ \varepsilon_\|(\omega)]$, and $\Real [ \varepsilon_\|(\omega)]$, versus energy ($\hslash\omega_g$), in eV.  The BSE spectra from this work (red) and from ref.~\citenum{TiO2GWVasp} (green) are based on $G_0W_0$ eigenvalues.  The experimental spectra (blue) are obtained from reflection spectra polarized perpendicular to the \textbf{(a,b)} $c$-axis or \textbf{(c,d)} $a$-axis by Kramers-Kronig transformation from Ref.~\citenum{TiO2AnataseReflectivityDielectricConstantExp}. 
}\label{fgr:AnataseEpsilon}
\end{figure}

Based on these \oldRev{QP }$G_0W_0$ levels, we obtain from the Bethe-Saltpeter equation the imaginary and real parts of the dielectric function of bulk A-TiO$_2$ for polarization perpendicular (ordinary) and parallel (extraordinary) to the tetragonal axis $c$ shown in Figure~\ref{fgr:AnataseEpsilon}.  These are comparable to the dielectric functions obtained from reflection spectra polarized perpendicular to the $a$ or $c$-axis at room temperature by Kramers-Kronig transformations\cite{TiO2AnataseReflectivityDielectricConstantExp}. Note that 86\% of the experimental reflectivity spectra polarized perpendicular to the $a$-axis is parallel to the $c$-axis\cite{TiO2AnataseReflectivityDielectricConstantExp}. Furthermore, our dielectric functions agree well with those obtained from BSE calculations within the Tamm-Dancoff approximation \cite{TiO2GWVasp}.  In particular, we obtain excellent agreement both in position and intensity for the first bright exciton at $\sim 4$~eV, which is perpendicular to the $c$-axis.  The lowest energy BSE $G_0W_0$ transition is at $3.73$~eV, about $0.12$~eV below the \oldRev{QP }PBE $G_0W_0$ indirect electronic gap of A-TiO$_2$\newRev{, as shown in Table~\ref{Eg:tbl}}.  This is significantly higher than the estimated optical band gap of $3.42$~eV  reported in ref.~\citenum{UrbachTailTiO2}\nocite{UrbachTailTiO2}. 
\nocite{TiO2GWVasp}\nocite{TiO2AnataseReflectivityDielectricConstantExp}

However, we tend to underestimate the real part of the dielectric function, shown in Figure~\ref{fgr:AnataseEpsilon}\textbf{(b,d)}. For example, as reported in Table~\ref{epsilonM:tbl}, the dielectric constant $\varepsilon_\infty =  \varepsilon(\omega=0)$ is underestimated by about 2 in our BSE calculations. This might be remedied by including a greater number of transitions within the BSE calculation.  

\begin{table}[!t]
\caption{\textrm{\bf Macroscopic Dielectric Constants $\boldsymbol{\varepsilon_{\infty}}$ \oldRev{Parallel}\newRev{Perpendicular} ($\boldsymbol{\perp}$) and Parallel ($\boldsymbol{\|}$) to the Tetragonal $\textit{c}$-axis of A-TiO$_{\text{2}}$ and R-TiO$_{\text{2}}$.}
}\label{epsilonM:tbl}
\begin{tabular}{ccllll}
\multicolumn{6}{>{\columncolor[gray]{0.9}}c}{ }\\[-3mm]
\multicolumn{1}{>{\columncolor[gray]{0.9}}c}{method} &
\multicolumn{1}{>{\columncolor[gray]{0.9}}c}{xc-functional} &
\multicolumn{2}{>{\columncolor[gray]{0.9}}c}{A-TiO$_2$} &
\multicolumn{2}{>{\columncolor[gray]{0.9}}c}{R-TiO$_2$} \\
\multicolumn{2}{>{\columncolor[gray]{0.9}}c}{ } 
 &\multicolumn{1}{>{\columncolor[gray]{0.9}}c}
{$\perp$} 
&\multicolumn{1}{>{\columncolor[gray]{0.9}}c}
{$\|$} 
&\multicolumn{1}{>{\columncolor[gray]{0.9}}c}
{$\perp$} 
&\multicolumn{1}{>{\columncolor[gray]{0.9}}c}
{$\|$} \\[1mm]
\multirow{6}{*}{RPA} 
& \multirow{2}{*}{LDA} & \multirow{2}{*}{7.18} & \multirow{2}{*}{6.81}
& 7.83$^a$ & 9.38$^a$ \\
& & & & 7.69$^b$ & 8.91$^b$ \\
& \multirow{2}{*}{PBE} &\multirow{2}{*}{7.06} & \multirow{2}{*}{6.60} & 7.61\oldRev{$^c$} & 9.09\oldRev{$^c$} \\
&  & & & 7.55$^b$ & 9.02$^b$ \\
& \multirow{2}{*}{HSE} &\multirow{2}{*}{4.91} & \multirow{2}{*}{4.83} & 5.21\oldRev{$^c$} & 6.09\oldRev{$^c$} \\
&  & & & 5.74$^b$ & 6.77$^b$ \\
\multirow{2}{*}{BSE-$G_0W_0$} & LDA & 4.17 & 3.45
& 5.60$^a$ & 7.11$^a$ \\

& PBE & 4.91\oldRev{$^d$}\newRev{$^c$} & 4.76\oldRev{$^d$}\newRev{$^c$} & 5.15\oldRev{$^d$}\newRev{$^c$} & 6.22\oldRev{$^d$}\newRev{$^c$}\\

 BSE-DFT & PBE+$\Delta$ & 5.12\oldRev{$^e$}\newRev{$^d$} & 4.98\oldRev{$^e$}\newRev{$^d$} & 5.71\oldRev{$^e$}\newRev{$^d$} & 7.33\oldRev{$^e$}\newRev{$^d$}\\
& & &
& 5.79\oldRev{$^h$}\newRev{$^e$} & 7.04\oldRev{$^h$}\newRev{$^e$}\\
\multicolumn{2}{c}{Experiment} 
& 5.73\oldRev{$^g$}\newRev{$^f$} & 5.64\oldRev{$^g$}\newRev{$^f$} 
& 5.88\oldRev{$^i$}\newRev{$^g$} & 7.14\oldRev{$^i$}\newRev{$^g$}\\
& &&
& 6.84\oldRev{$^f$}\newRev{$^h$} & 8.43\oldRev{$^f$}\newRev{$^h$}\\[2mm]
\multicolumn{6}{p{0.95\columnwidth}}{\tiny$^a$Ref.~\citenum{MiganiLong}. $^b$Ref.~\citenum{TiO2DielectricConstantLDA}\nocite{TiO2DielectricConstantLDA}. 
\oldRev{$^c$Obtained using the computational parameters reported in Ref.~\citenum{MiganiLong}. }
\oldRev{$^d$}\newRev{$^c$}Ref.~\citenum{TiO2GWVasp}. \oldRev{$^e$}\newRev{$^d$}Ref.~\citenum{TiO2OpticalPropertiesBSEDFT+U}\nocite{TiO2OpticalPropertiesBSEDFT+U}. \oldRev{$^f$Ref.~\citenum{TiO2DielectricConstantExp}\nocite{TiO2DielectricConstantExp}. \oldRev{$^g$}\newRev{$^f$}Ref.~\citenum{TiO2AnataseReflectivityDielectricConstantExp}\nocite{TiO2AnataseReflectivityDielectricConstantExp}.}  \oldRev{$^h$}\newRev{$^e$}Ref.~\citenum{OpticalTiO2}\nocite{OpticalTiO2}. \newRev{$^f$Ref.~\citenum{TiO2AnataseReflectivityDielectricConstantExp}\nocite{TiO2AnataseReflectivityDielectricConstantExp}. }\oldRev{$^i$}\newRev{$^g$}Ref.~\citenum{SPIE}\nocite{SPIE}. \newRev{$^h$Ref.~\citenum{TiO2DielectricConstantExp}\nocite{TiO2DielectricConstantExp}.}
}
\end{tabular}
{\color{StyleColor}{\rule{\columnwidth}{1.0pt}}}
\end{table}

In any case, such computationally demanding calculations are beyond the scope of the present work.  Overall, the agreement obtained for the BSE dielectric function based on \oldRev{QP }$G_0W_0$ eigenenergies demonstrates the robustness of the parameters we will use to calculate the \oldRev{QP }$G_0W_0$ PDOS for H$_2$O.

Figure~\ref{fgr:CleanSurfaceSchematic} depicts schematically the clean and stoichiometric A-TiO$_2$(101) surface. For the clean surface, there are 
two Ti coordinately unsaturated sites (Ti$_{\textit{cus}}$) and two bridging
O atoms (O$_{\textit{br}}$) in each unit cell. 

\begin{figure*}
\includegraphics[width=\textwidth]{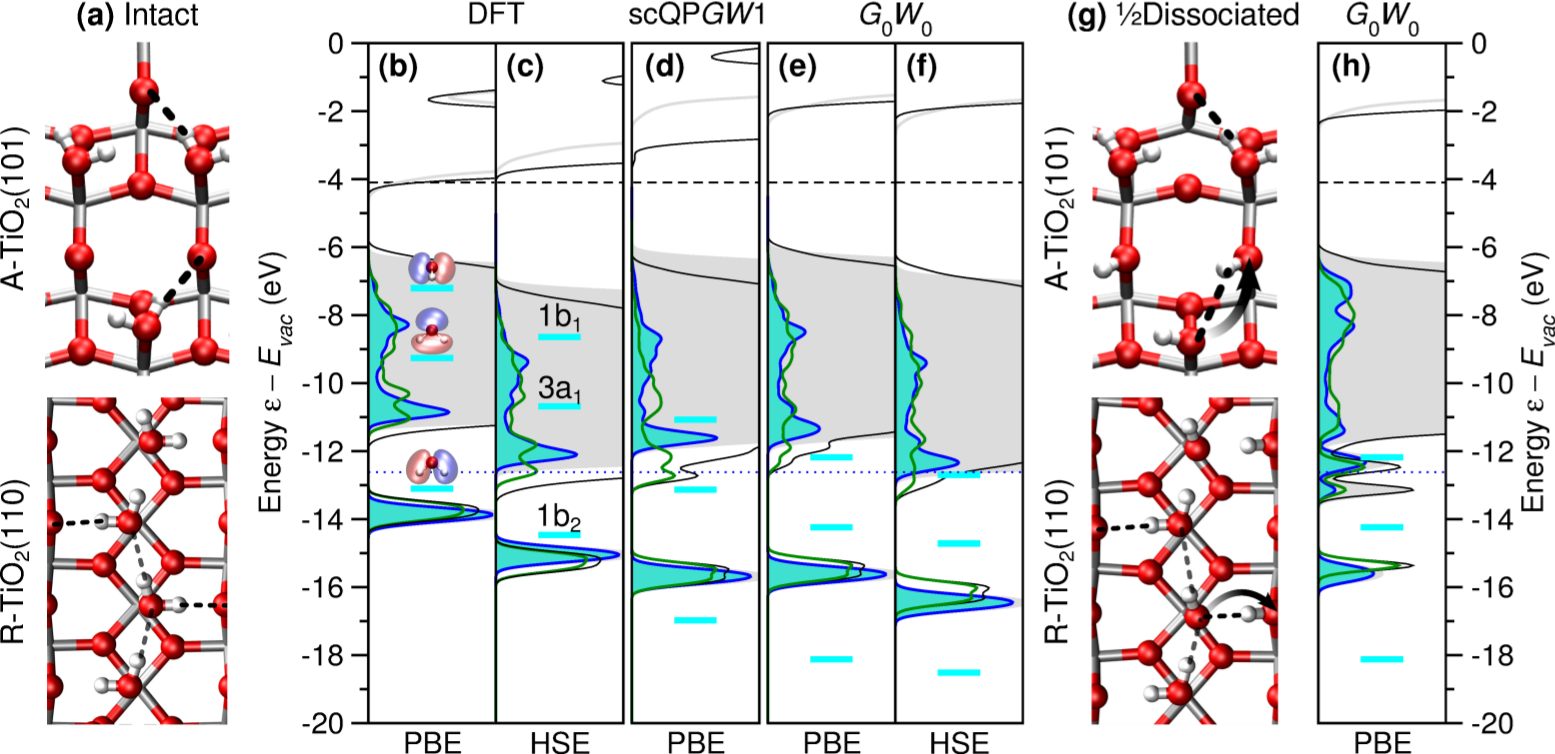}
\caption[H$_2$O@Ti$_{\textit{cus}}$]{\textbf{(a-f)} 1ML intact and \textbf{(g,h)} \sfrac{1}{2} dissociated H$_{\text{2}}$O adsorbed on coordinately unsaturated Ti sites (H$_{\text{2}}$O@Ti$_{\textit{cus}}$). A-TiO$_2$(101)/R-TiO$_2$(110)\cite{MiganiH2O} total (grey/black) and H$_2$O projected (blue/green) DOS  computed with \textbf{(b,c)} DFT, \textbf{(d)} scQP$GW$1, and \textbf{(e,f,h)} $G_0W_0$ using the \textbf{(b,d,e,h)} generalized gradient approximation (PBE)\cite{PBE} and \textbf{(c,f)} range-separated hybrid (HSE)\cite{HSE06} for the xc-functional. Filling denotes occupation for A-TiO$_{2}$(101). Energies are relative to the vacuum level, $E_{\textit{vac}}$.  The measured $\varepsilon_{\textrm{CBM}}$\cite{SprikH2OAlignment} (black dashed line), measured and coupled-cluster (CCSD(T)) H$_2$O gas phase ionization potentials \textit{IP}\cite{GWMoleculesMarquesJCTC2013} (blue dotted line), and for each level of theory the calculated gas phase 1b$_1$, 3a$_1$, and 1b$_2$ H$_2$O levels\cite{MiganiH2O} (marked in cyan) are provided. \newRev{Note that the fully symmetric 2a$_1$ H$_2$O levels lie below -20~eV. }\textbf{(a,g)} Charge transfer of about $-0.4e$ accompanying deprotonation is represented by arrows, while intermolecular (gray) and interfacial (black) hydrogen bonds are denoted by dotted lines. Ti, O, and H atoms are depicted in silver, red, and white, respectively.
}\label{fgr:TicusEvac}
{\color{StyleColor}{\rule{\textwidth}{1.0pt}}}
\end{figure*}

PBE $G_0W_0$ places the $\textit{IP}$ for A-TiO$_2$(101) at 7.15~eV, which is 0.14 eV below that of R-TiO$_2$(110)\cite{MiganiLong}.  This relative ordering is consistent with, albeit significantly smaller than, that measured with XPS for the A-TiO$_2$--RuO$_2$--R-TiO$_2$ interface of $0.7\pm 0.1$~eV\cite{ExpAandRTiO2alignment}. This ordering also agrees with the 0.47~eV difference in \textit{IP} calculated using a hybrid quantum-mechanical/molecular mechanical embedding technique\cite{BandAlignmentR-A}. In these calculations the $\textit{IP}$ was obtained from the total energy difference upon removal of an electron from the neutral A-TiO$_2$ and R-TiO$_2$ embedded cluster models.  Our relative ordering is also consistent with that obtained from \oldRev{Kohn-Sham}\newRev{KS} eigenvalues using the B3LYP xc-functional of 8.24\cite{HarrisonPRB2014} and 8.6~eV\cite{SprikChemCatChem} for A-TiO$_2$(101) and R-TiO$_2$(110), respectively.  This qualitative agreement provides further support for the reliability of our \oldRev{QP }$G_0W_0$ approach.

\subsection{H$_{\text{2}}$O Intact and \sfrac{1}{2} Dissociated on Coordinately Unsaturated Ti Sites}\label{Sect:Intact}

For coverages up to 1ML, H$_2$O adsorbs molecularly on the A-TiO$_2$(101) surface, with O bonding to Ti$_{\textit{cus}}$ and \newRev{one H forming an }interfacial hydrogen bond\oldRev{s} with \oldRev{the }O$_{\textit{br}}$\oldRev{ atoms}\cite{Selloni1998219,SelloniWaterReview2010,DieboldSelloniNatMater2009,PhysRevLett.81.2954,DieboldJPCB2003}, as shown in Figure \ref{fgr:TicusEvac}\textbf{(a)}. On R-TiO$_2$(110), the distance between the nearest neighboring Ti$_{\textit{cus}}$ sites is shorter, allowing \newRev{additional }intermolecular hydrogen bonds to form along the [001] direction \cite{KennethJordanWaterChain,KimmelJPCL2012,MichaelidesDynamics,JinH2O}, as show in Figure \ref{fgr:TicusEvac}\textbf{(a)}.

The QP level alignment relative to the vacuum level $E_{\textit{vac}}$ for 1ML of H$_2$O adsorbed intact on A-TiO$_2$(101) and R-TiO$_2$(110)\cite{MiganiH2O} are shown in Figure~\ref{fgr:TicusEvac}\textbf{(b-f)}.  These are compared to the 1b$_2$, 3a$_1$, and 1b$_1$ levels' absolute alignment for gas phase H$_2$O\cite{MiganiH2O}. Specifically, we analyze the dependence of the H$_2$O PDOS on the methodology: PBE DFT, HSE DFT, PBE scQP$GW\oldRev{_1}\newRev{1}$, 
PBE $G_0W_0$, and HSE $G_0W_0$.  

As was previously found for the H$_2$O--R-TiO$_2$(110) interface, the $\textit{IP}$ for H$_2$O--A-TiO$_2$(101) is ordered according to the method's description of the screening, $\varepsilon_\infty^{-1}$  \cite{MiganiH2O}.  \oldRev{That is}\newRev{As shown in Table~\ref{IP:tbl}}, the $\textit{IP}$ is ordered as PBE $G_0W_0$ (6.3~eV) $\approx$ PBE scQP$GW1$ (6.3~eV) $\sim$ PBE DFT (6.4~eV) $<$ HSE $G_0W_0$ (6.9~eV) $<$ HSE DFT (7.2~eV).   

\begin{table}
\caption{\newRev{\textrm{\bf Ionization Potentials \textit{IP} in eV of 1ML H$_{\textit{2}}$O@Ti$_{\textit{cus}}$ on A-TiO$_{\text{2}}$(101) and R-TiO$_{\text{2}}$(110).}}
}\label{IP:tbl}
\begin{tabular}{cccc}
\multicolumn{4}{>{\columncolor[gray]{0.9}}c}{ }\\[-3mm]
\multicolumn{1}{>{\columncolor[gray]{0.9}}c}{method} &
\multicolumn{1}{>{\columncolor[gray]{0.9}}c}{xc-functional} &
\multicolumn{2}{>{\columncolor[gray]{0.9}}c}{H$_2$O@Ti$_{\textit{cus}}$}\\
\multicolumn{2}{>{\columncolor[gray]{0.9}}c}{} & 
\multicolumn{1}{>{\columncolor[gray]{0.9}}c}{A-TiO$_2$(101)} &
\multicolumn{1}{>{\columncolor[gray]{0.9}}c}{R-TiO$_2$(110)} \\[1mm]
\multirow{2}{*}{DFT} & PBE & 6.4 & 6.2$^a$\\
& HSE & 7.2 & 7.3$^a$ \\
scQP$GW1$ & PBE & 6.3 & 6.6$^a$\\
\multirow{2}{*}{$G_0W_0$} & PBE & 6.3 & 6.0$^a$\\
& HSE & 6.9 & 6.5$^a$\\
\multicolumn{4}{p{0.95\columnwidth}}{\tiny$^a$Ref.~\citenum{MiganiH2O}}
\end{tabular}
{\color{StyleColor}{\rule{\columnwidth}{1.0pt}}}
\end{table}

Note that the CBM and VBM relative to $E_{\textit{vac}}$ from PBE scQP$GW1$ and PBE $G_0W_0$ are essentially the same for H$_2$O--A-TiO$_2$(101), but are significantly lower for H$_2$O--R-TiO$_2$(110). This is because the dielectric constant employed in both single-shot PBE scQP$GW1$ and PBE $G_0W_0$ are those obtained from PBE DFT, whereas when the QP self energies are ``mixed'' with the DFT xc-potential in each cycle, as for H$_2$O--R-TiO$_2$(110), the scQP$GW1$ dielectric constant is significantly reduced relative to PBE DFT.  This demonstrates that without mixing of the self energy, for the QP PDOS the PBE scQP$GW1$ procedure provides no advantage over PBE $G_0W_0$, as predicted in ref.~\citenum{MiganiH2O}.

Generally, the highest H$_2$O PDOS peaks, $\epeak$, follow the same ordering as the $\textit{IP}$s.  This suggests that $\epeak$ is pinned to the VBM of the H$_2$O--A-TiO$_2$(101) interface.  This is also the case for 1ML intact H$_2$O@Ti$_{\textit{cus}}$ on R-TiO$_2$(110)\cite{MiganiH2O}.  However, this ordering of the $\textit{IP}$s is completely different from that found for gas phase H$_2$O\cite{MiganiH2O}.  In this case, the $\textit{IP}$ is the energy needed to remove one electron from the H$_2$O 1b$_1$ level.  Here, the $\textit{IP}$s increase with decreasing screening within the methodology until $\varepsilon_\infty \sim 1$\cite{MiganiH2O}.

\begin{figure*}[!t]
\includegraphics[width=\textwidth]{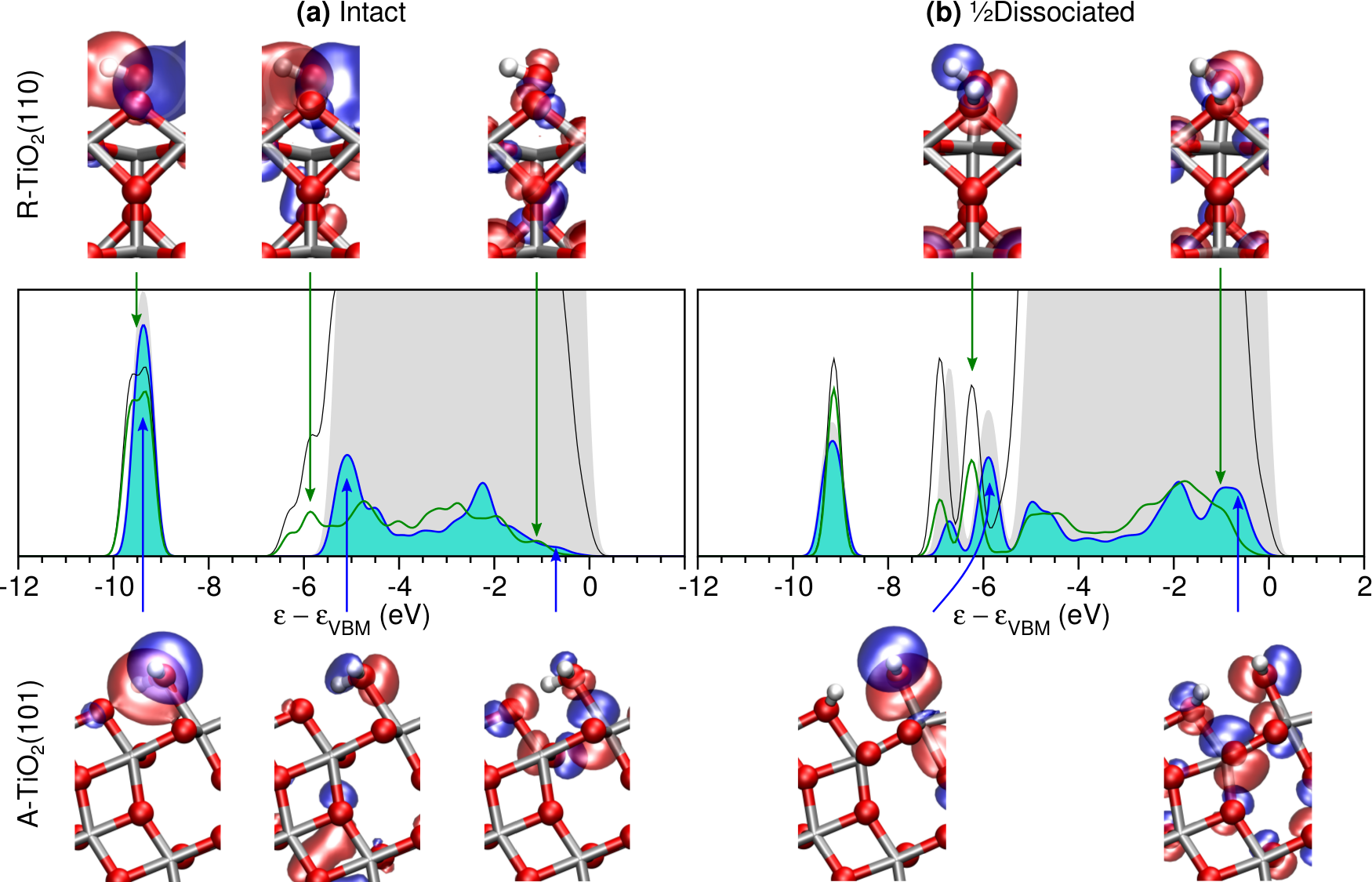}
\caption[H$_2$O@Ti$_{\textit{cus}}$]{\textbf{(a)} 1ML intact and \textbf{(b)} \sfrac{1}{2} dissociated H$_{\text{2}}$O adsorbed on coordinately unsaturated Ti sites (H$_{\text{2}}$O@Ti$_{\textit{cus}}$). Total (grey/black) and H$_2$O projected (blue/green) \oldRev{QP }$G_0W_0$ DOS on anatase TiO$_2$(101)/rutile TiO$_2$(110) surfaces and selected orbitals are shown schematically below/above. Energies are relative to the valence band maximum, $\eVBM$. Ti, O, and H atoms are depicted in silver, red, and white, respectively.
}\label{fgr:TicusVBM}
{\color{StyleColor}{\rule{\textwidth}{1.0pt}}}
\end{figure*}

However, for 1ML intact H$_2$O@Ti$_{\textit{cus}}$, the relative alignment of the A-TiO$_2$(101) and R-TiO$_2$(110) VBMs differs qualitatively with methodology.  The $\textit{IP}$s obtained from PBE DFT and PBE $G_0W_0$ are higher ($\sim 0.2$ and $0.3$~eV) for A-TiO$_2$(101) than for R-TiO$_2$(110).  However, the $IP$ from HSE DFT is lower ($\sim -0.1$~eV) for A-TiO$_2$(101) than R-TiO$_2$(110), while the opposite is true for HSE $G_0W_0$ ($\sim 0.4$~eV).  Thus, independently of the xc-functional employed, \oldRev{QP }$G_0W_0$ yields higher $\textit{IP}$s for 1ML intact H$_2$O@Ti$_{\textit{cus}}$ on A-TiO$_2$(101) than on R-TiO$_2$(110).  This is contrary to our findings for the clean A-TiO$_2$(101) and R-TiO$_2$(110) surface, and suggests that H$_2$O adsorption inverts the relative positions of the A-TiO$_2$(101) and R-TiO$_2$(110) VBMs.  

Although we find the position of the lower edge of the valence band is only weakly affected by adsorbing H$_2$O on either A-TiO$_2$(101) or R-TiO$_2$(110), the VBM is shifted up by about 1 eV in both cases.  This is consistent with the experimentally observed change in work function for the liquid water--R-TiO$_2$(110) interface\cite{MiganiLong,TiO2WorkFunction30,Onishi198833,C0CP02835E,Petek2PPH2O,SprikH2OAlignment,MiganiH2O,PetekScienceH2O}.

The reordering of the HSE DFT and $G_0W_0$ $\textit{IP}$s  for 1ML intact H$_2$O on A-TiO$_2$(101) and R-TiO$_2$(110) may be attributed to the greater difference between the constant screening built into HSE DFT \cite{Marques} and the screening of rutile compared to anatase.  
Essentially, the fraction of the Hartree-Fock exact-exchange which is incorporated within HSE, $\alpha = 0.25$, acts as an effective inverse dielectric constant within the system, $\alpha \sim \varepsilon_\infty^{-1}$ \cite{Marques}.  As a result, for materials with $\varepsilon_\infty \approx 4$, HSE DFT and $G_0W_0$ should provide similar descriptions of the screening \cite{Marques}.  From Table~\ref{epsilonM:tbl}, we see that the RPA, BSE, and measured $\varepsilon_\infty$ agree qualitatively and are consistently lower and closer to the HSE DFT effective dielectric constant of $\varepsilon_\infty \sim 4$ for A-TiO$_2$ compared to R-TiO$_2$.  For this reason, as shown in Figure~\ref{fgr:TicusEvac}, the difference between HSE DFT and $G_0W_0$ \textit{IP}s is larger for R-TiO$_2$ than A-TiO$_2$, resulting in their relative reordering at the $G_0W_0$ level.
This demonstrates the important role played by the screening in describing the relative alignment of anatase and rutile polymorphs.  

Overall the H$_2$O QP PDOS for 1ML intact H$_2$O@Ti$_{\textit{cus}}$ is similar for the A-TiO$_2$(101) and R-TiO$_2$(110) surfaces.  In particular, the most strongly bound 1b$_2$ peaks and the upper edges of the H$_2$O PDOS spectra have similar energies for the two polymorphs over all five levels of theory (\emph{cf.} Figure~\ref{fgr:TicusEvac}\textbf{(b-f)}).

On A-TiO$_2$(101), the 1ML intact H$_2$O QP PDOS generally consists of three distinct peaks, which have clear contributions from molecular 1b$_2$, 3a$_1$ and 1b$_1$ levels (\emph{cf.} Figure~\ref{fgr:TicusEvac}\textbf{(b)} and Figure~\ref{fgr:TicusVBM}\textbf{(a)}).  This is in contrast to R-TiO$_2$(110), where the H$_2$O QP PDOS consists of many more peaks, with a greater hybridization at 1ML compared to \sfrac{1}{2}ML coverage on the R-TiO$_2$(110) substrate.\cite{MiganiH2O}  This may be attributed to stronger intermolecular interactions on R-TiO$_2$(110) due to its shorter Ti$_{\textit{cus}}$ nearest neighbor separations ($d[\textrm{Ti}_{\textrm{cus}}-\textrm{Ti}_{\textrm{cus}}] \approx 2.96$~\AA) versus A-TiO$_2$(101) ($d[\textrm{Ti}_{\textrm{cus}}-\textrm{Ti}_{\textrm{cus}}] \approx 3.78$~\AA).  This leads to intermolecular bonding and antibonding levels, which may further hybridize with the substrate.\cite{MiganiH2O}
For example, as shown in Figure~\ref{fgr:TicusEvac}, the bottom edge of the 3a$_1$ peak for A-TiO$_2$(101) is higher than that of R-TiO$_2$(110).  This is because on R-TiO$_2$(110) the 3a$_1$ levels of neighbouring molecules hybridize to form intermolecular bonding and antibonding combinations\cite{MiganiH2O}. These give rise to separate peaks below and above the bottom edge of the R-TiO$_2$(110) valence band.  As a result, the QP H$_2$O PDOS for 1ML intact H$_2$O@Ti$_{\textit{cus}}$ on R-TiO$_2$(110) has the 3a$_1$ intermolecular bonding level below the bottom of the valence band, while for A-TiO$_2$(101), the 3a$_1$ level is completely within the substrate's valence band.

\begin{figure*}
\includegraphics[width=\textwidth]{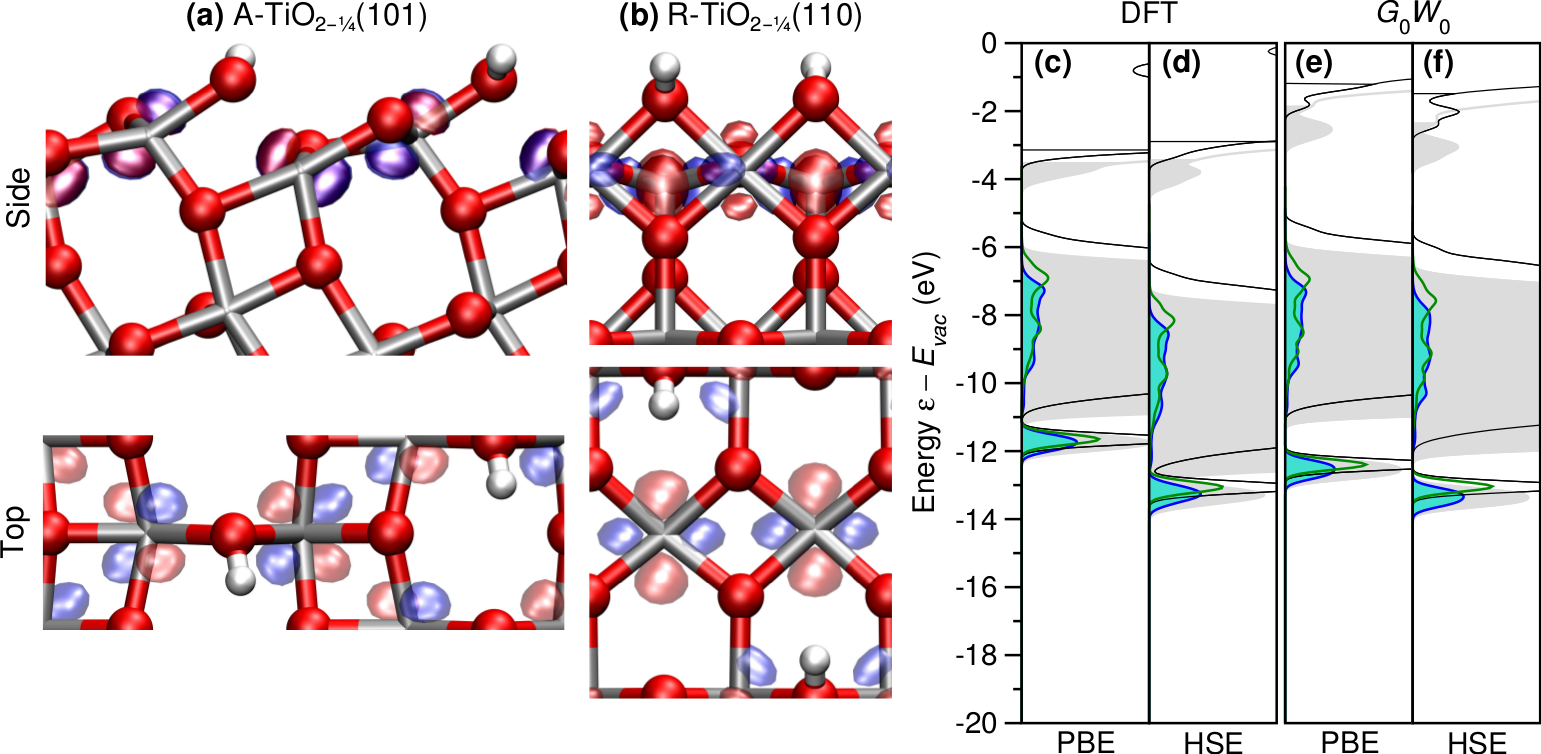}
\caption[H$_2$O@Ti$_{\textit{cus}}$]{
\sfrac{1}{2}ML H$_{\text{2}}$O dissociated on bridging O vacancies (H$_2$O@O$_{\textit{br}}^{\textit{vac}}$) of defective \textbf{(a/b)} A-TiO$_{2-\text{\sfrac{1}{4}}}$(101)/R-TiO$_{2-\text{\sfrac{1}{4}}}$(110)\cite{MiganiH2O} surfaces with \sfrac{1}{2}ML of O$_{\textit{br}}^{\textit{vac}}$. Total (grey/black) and H$_2$O projected (blue/green) DOS  computed with \textbf{(c,d)} DFT and \textbf{(e,f)} $G_0W_0$ using the \textbf{(c,e)} generalized gradient approximation (PBE)\cite{PBE} and \textbf{(d,f)} range-separated hybrid (HSE)\cite{HSE06} for the xc-functional. Filling denotes occupation for A-TiO$_{2-\text{\sfrac{1}{4}}}$(101).  Horizontal black lines denote Fermi levels $\eF$ for R-TiO$_{2-\text{\sfrac{1}{4}}}$(110).  Energies are relative to the vacuum level, $E_{\textit{vac}}$.  Ti, O, and H atoms are depicted in silver, red, and white, respectively.
}\label{fgr:ObrvacEvac}
{\color{StyleColor}{\rule{\textwidth}{1.0pt}}}
\end{figure*}

Figure~\ref{fgr:TicusEvac}\textbf{(g)} shows the structures of \sfrac{1}{2} dissociated H$_2$O@Ti$_{\textit{cus}}$ on A-TiO$_2$(101) and R-TiO$_2$(110). In both cases, one proton from H$_2$O@Ti$_{\textit{cus}}$ is transferred to the adjacent O$_{\textit{br}}$.  This results in two distinct OH groups: HO@Ti$_{\textit{cus}}$ and O$_{\textit{br}}$H.  This process is accompanied by a $-0.4e$ charge transfer from HO@Ti$_{\textit{cus}}$ to O$_{\textit{br}}$H, as depicted schematically in Figure~\ref{fgr:TicusEvac}\textbf{(g)}.

Although the resulting PBE $G_0W_0$ QP DOS shown in Figure~\ref{fgr:TicusEvac}\textbf{(e)} and \textbf{(h)} are generally similar, there are important differences which are related to the H$_2$O@Ti$_{\textit{cus}}$ dissociation.  For the dissociated H$_2$O molecule, the 1b$_2$ peak is replaced by separate HO@Ti$_{\textit{cus}}$  \oldRev{($\sim-12.5$~eV) }and O$_{\textit{br}}$H \oldRev{($\sim-13$~eV) }peaks \newRev{at about $-12.5$ and $-13$~eV below $E_{\textit{vac}}$ (see Figure~\ref{fgr:TicusEvac}\textbf{(h)}),} with O--H $\sigma$ character  on both A-TiO$_2$(101) and R-TiO$_2$(110)\cite{MiganiH2O} (see Figure~\ref{fgr:TicusVBM}\textbf{(b)}).  These peaks are at such similar energies on both A-TiO$_2$(101) and R-TiO$_2$(110) because they are well separated from the bottom edge of the TiO$_2$ valence band.

As mentioned above, the three distinct peaks in the H$_2$O PDOS on both A-TiO$_2$(101) and R-TiO$_2$(110) are associated with the 1b$_2$, 3a$_1$, and 1b$_1$ gas phase H$_2$O levels.  This is clearly seen by comparing the molecular components of the orbitals depicted for 1ML intact and \sfrac{1}{2} dissociated H$_2$O@Ti$_{\textit{cus}}$ on both A-TiO$_2$(101) and R-TiO$_2$(110) in Figure~\ref{fgr:TicusVBM} with the gas phase H$_2$O levels depicted in Figure~\ref{fgr:TicusEvac}.  There is significantly greater hybridization between the molecular levels on R-TiO$_2$(110) compared to A-TiO$_2$(101).  Specifically, on R-TiO$_2$(110) there are obvious bonding and antibonding combinations of the 1b$_2$ levels and 3a$_1$ levels between neighbouring H$_2$O molecules.\cite{MiganiH2O}  Such intermolecular hybridization does not occur for A-TiO$_2$(101), as the molecules are too far apart.  

More importantly, $\epeak$ is shifted to higher energy upon dissociation, with a greater shift for A-TiO$_2$(101) versus R-TiO$_2$(110).  To explain these differences, and their potential impact on the interfaces' photocatalytic activity, one should compare the level alignment relative to the VBM.  In so doing, one can directly compare the relative propensity for photogenerated hole transfer from the substrate's  VBM to the molecular HOMO for A-TiO$_2$(101) and R-TiO$_2$(110).  

In Figure~\ref{fgr:TicusVBM} we provide the level alignment relative to the VBM for \textbf{(a)} intact and \textbf{(b)} \sfrac{1}{2} dissociated H$_2$O on A-TiO$_2$(101) and R-TiO$_2$(110). The level alignment shown in Figure~\ref{fgr:TicusVBM} suggests that (1) hole trapping is more favored on A-TiO$_2$(101) than R-TiO$_2$(110) and (2) HO@Ti$_{\textit{cus}}$ is more photocatalytically active than intact H$_2$O@Ti$_{\textit{cus}}$.  This is based on the following observations: 
(1) $\epeak$ is about 0.5 eV higher in energy for A-TiO$_2$(101) than R-TiO$_2$(110); 
(2) $\epeak$ is about 0.1 eV closer to the VBM for HO@Ti$_{\textit{cus}}$ than for intact H$_2$O@Ti$_{\textit{cus}}$; 
(3) the PDOS for HO@Ti$_{\textit{cus}}$ at $\epeak$ is an order of magnitude greater than for intact H$_2$O@Ti$_{\textit{cus}}$; 
and (4) the HOMO is more localized on the molecule for HO@Ti$_{\textit{cus}}$  than for intact H$_2$O@Ti$_{\textit{cus}}$.

These conclusions are reinforced by analyzing the HOMOs at $\Gamma$ shown in Figure~\ref{fgr:TicusVBM}.  Here, one clearly sees that the HOMOs have greater weight on the molecule for HO@Ti$_{\textit{cus}}$ than intact H$_2$O@Ti$_{\textit{cus}}$.  This should promote hole trapping on HO@Ti$_{\textit{cus}}$.
Although there is only a small (0.1 eV) energy difference between the HOMO for \sfrac{1}{2} dissociated and intact H$_2$O@Ti$_{\textit{cus}}$, the latter level is not photocatalytically relevant for hole trapping \newRev{on the molecule}.  This is because it is a lone-pair orbital that datively bonds to Ti$_{\textit{cus}}$.  For this reason, if an electron were extracted from this level, one would \newRev{instead} expect \newRev{the hole to remain on the surface, and} H$_2$O to desorb from Ti$_{\textit{cus}}$.  \newRev{This agrees with previous studies of the liquid H$_2$O--A-TiO$_2$(101) interface, which found that localizing the hole on intact H$_2$O is inherently unstable, and leads to deprotonation\cite{SelloniPCETJACS2013}.  Instead, the hole localizes on 3-fold coordinated surface O (O$_{3\textit{fold}}$) atoms\cite{SelloniPCETJACS2013}.}

In contrast to the intact H$_2$O@Ti$_{\textit{cus}}$ HOMOs, the HOMOs for HO@Ti$_{\textit{cus}}$ on A-TiO$_2$(101) and R-TiO$_2$(110) are the photocatalytically active levels for hole-trapping.  Indeed, they have the same character as the hole trapping levels reported in the literature for A-TiO$_2$(101)\cite{SelloniPCETJACS2013} and R-TiO$_2$(110)\cite{SprikSchematic}.  In particular, they have both \oldRev{3-fold coordinate surface O (}O$_{3\textit{fold}}$\oldRev{)} 2p$_\pi$ \cite{DuncanTiO2} and OH 2p character.  While in the case of HO@Ti$_{\textit{cus}}$, this orbital is doubly occupied, in the trapped hole structures of refs.~\citenum{SelloniPCETJACS2013} and \citenum{SprikSchematic}, the OH groups are bent towards the surface, with the hole shared between O$_{3\textit{fold}}$ 2p$_\pi$ and OH 2p  orbitals.  

\newRev{This clearly demonstrates that a HOMO initially below the VBM can, upon light absorption and subsequent nuclear relaxation, evolve into a hole trapping level of the interface.  This justifies our use of ground state level alignment for comparing photocatalytic
activity among H$_2$O--TiO$_2$ interfaces.  }

Although hole trapping has been documented for both A-TiO$_2$(101) \cite{SelloniPCETJACS2013} and R-TiO$_2$(110) \cite{SprikSchematic,Tritsaris-KaxirasJPCC2014}, the more favorable ground state HOMO level alignment for A-TiO$_2$(101) may explain why the anatase polymorph shows higher photocatalytic activity than the rutile polymorph \cite{GratzelJACS1996,A-TiO2101WaterSplitting,AvsRTiO2ActivityThinFilmsSciRep2014}.

\subsection{H$_{\text{2}}$O Dissociated on Bridging O Vacancies}\label{Sect:Dissociated}

For R-TiO$_2$(110), the most stable O vacancies are at surface O$_{\textit{br}}$ sites, i.e., O$_{\textit{br}}^{\textit{vac}}$.  These sites mediate H$_2$O dissociation on R-TiO$_2$(110) \cite{ThorntonH2ODissTiO2110,NorskovVacanciesPRL2001}.  For A-TiO$_2$(101), the most stable O vacancies are subsurface \cite{Setvin30082013,H2OonSubsurfaceDefects}. However, after H$_2$O adsorption, these subsurface vacancies migrate to the surface and are filled by H$_2$O, i.e., H$_2$O@O$_{\textit{br}}^{\textit{vac}}$, which subsequently dissociates \newRev{to form 2HO$_{\textit{br}}$}\cite{YadongPRLH2OSubsurface,H2OonSubsurfaceDefects}.  This results in a structure equivalent to H on a stoichiometric A-TiO$_2$(101) surface \cite{YadongPRLH2OSubsurface,H2OonSubsurfaceDefects,C2CP42288C}.   
For this reason, we consider a \sfrac{1}{2}ML coverage of H$_2$O adsorbed dissociatively on O$_{\textit{br}}^{\textit{vac}}$ sites (H$_2$O@O$_{\textit{br}}^{\textit{vac}}$) of a defective A-TiO$_{2-\text{\sfrac{1}{4}}}$(101) or R-TiO$_{2-\text{\sfrac{1}{4}}}$(110) \cite{MiganiH2O} surface consisting of \sfrac{1}{2}ML of O$_{\textit{br}}^{\textit{vac}}$, shown schematically in Figure~\ref{fgr:ObrvacEvac}.  This is equivalent to 1ML of H adsorbed on O$_{\textit{br}}$ (H@O$_{\textit{br}}$) of a stoichiometric A-TiO$_2$(101) or R-TiO$_2$(110) surface. 

These hydroxylated structures have occupied Ti $3d$ levels which are associated with reduced Ti$^{3+}$ atoms.  The excess electrons introduce $n$-type doping.  These occupied Ti$^{3+}$ 3$d$ levels give rise to the charge density just below the Fermi level, $\eF$,  in the DOS shown in Figure~\ref{fgr:ObrvacEvac}\textbf{(c-f)}\cite{PetekScienceH2O}.   

The difference in spatial and energetic localization of the Ti$^{3+}$ $3d$ levels between O defective A-TiO$_2$(101) and R-TiO$_2$(110) has been recently probed via STM \cite{PhysRevLett.113.086402}.  For O$_{\textit{br}}^{\textit{vac}}$@A-TiO$_2$(101)  at 6~K, the excess electrons are strictly localized next to  O$_{\textit{br}}^{\textit{vac}}$\cite{PhysRevLett.113.086402}, while for O$_{\textit{br}}^{\textit{vac}}$@R-TiO$_2$(110) at 78~K, the excess electrons are not confined next to O$_{\textit{br}}^{\textit{vac}}$\cite{PhysRevLett.113.086402,JinZhaoJCP2009STM}.  Instead, the excess electrons in  O$_{\textit{br}}^{\textit{vac}}$@R-TiO$_2$(110) may occupy $3d$ levels of surface Ti$_{\textit{cus}}$ or subsurface Ti atoms.

We find for \sfrac{1}{2}ML H$_2$O@O$_{\textit{br}}^{\textit{vac}}$ on both A-TiO$_{2-\textrm{\sfrac{1}{4}}}$(101) and R-TiO$_{2-\textrm{\sfrac{1}{4}}}$(110), the highest energy occupied Ti$^{3+}$ $3d$ levels\cite{DuncanTiO2} are mostly on surface Ti atoms, as shown in Figure~\ref{fgr:ObrvacEvac}\textbf{(a,b)}. These predominantly Ti $3d_{x^2-y^2}$ levels\cite{DuncanTiO2} are bonding along the [010] and [001] directions for A-TiO$_{2-\textrm{\sfrac{1}{4}}}$(101) and R-TiO$_{2-\textrm{\sfrac{1}{4}}}$(110), respectively.  Furthermore, for H$_2$O@O$_{\textit{br}}^{\textit{vac}}$ on A-TiO$_{2-\textrm{\sfrac{1}{4}}}$(101), the level occupies HO$_{\textit{br}}$'s  nearest neighbor Ti atoms.  For H$_2$O@O$_{\textit{br}}^{\textit{vac}}$ on R-TiO$_{2-\textrm{\sfrac{1}{4}}}$(110), this level also has weight on the next next nearest neighbour Ti$_{\textit{cus}}$ atoms. Additionally, there are higher energy occupied Ti $3d$ levels on subsurface Ti atoms.

In PBE DFT, the occupied Ti $3d$ levels form a shoulder at the bottom edge of the conduction band for H$_2$O@O$_{\textit{br}}^{\textit{vac}}$ on A-TiO$_{2-\textrm{\sfrac{1}{4}}}$(101), whereas on R-TiO$_{2-\textrm{\sfrac{1}{4}}}$(110) they do not even form a shoulder, as shown in Figure~\ref{fgr:ObrvacEvac}\textbf{(c)}.  
The degree of energetic localization of the Ti$^{3+}$ $3d$ levels, and their energy $\eTid$ below $\eF$, increases with the level of theory from PBE DFT $<$ HSE DFT ($\eTid \sim 0.6, 0.4$~eV) $<$ PBE $G_0W_0$ ($\eTid \sim 0.7, 0.6$~eV) $<$ HSE $G_0W_0$ ($\eTid \sim 1.0, 0.9$~eV) for H$_2$O@O$_{\textit{br}}^{\textit{vac}}$ on  A-TiO$_{2-\textrm{\sfrac{1}{4}}}$(101)/R-TiO$_{2-\textrm{\sfrac{1}{4}}}$(110), and is generally higher (0.1~eV) for 
A-TiO$_{2-\textrm{\sfrac{1}{4}}}$(101) than  R-TiO$_{2-\textrm{\sfrac{1}{4}}}$(110)\newRev{, as shown in Table~\ref{Ti3+:tbl}}.    This is consistent with
 the $\eTid \sim 1$~eV measured for O defective A-TiO$_2$(101) and R-TiO$_2$(110) and HO$_{\textit{br}}$@R-TiO$_2$(110) by scanning tunneling spectroscopy (STS)\cite{Papageorgiou09022010,PhysRevLett.113.086402,JinZhaoJCP2009STM}, \oldRev{and }photoemission electron spectroscopy (PES)\newRev{,} \cite{PESXASRutileAnatasePRB2007,DeSegovia} \newRev{and two photon photoemission spectroscopy (2PP)}.\newRev{\cite{PetekScienceH2O,PetekPRB2015}} However, a full treatment of Ti $3d$ defect levels, e.g., due to interstitial Ti atoms, also requires the inclusion of electron-phonon interactions \cite{JinZhaoJCP2009STM,Ti3dDefectStatesPolaritonicEffects1}.

\begin{table}
\vspace{-12.5mm}\noindent{\color{StyleColor}{\rule{\columnwidth}{1.0pt}}}
\vspace{8mm}
\caption{\newRev{\textrm{\bf Occupied Ti$\boldsymbol{^{3+}}$ $\boldsymbol{3d}$ Level Energies $\boldsymbol{\varepsilon_ {\text{Ti}^{3+}}}$ in eV Below the Fermi Level $\boldsymbol{\varepsilon_{\text{F}}}$ for \sfrac{1}{2}ML Dissociated H$_{\text{2}}$O@O$_{\textit{br}}^{\textit{vac}}$ on A-TiO$\boldsymbol{_{\text{2}-\text{\sfrac{1}{4}}}}$(101) and R-TiO$\boldsymbol{_{\text{2}-\text{\sfrac{1}{4}}}}$(110).}}
}\label{Ti3+:tbl}
\begin{tabular}{cccc}
\multicolumn{4}{>{\columncolor[gray]{0.9}}c}{ }\\[-3mm]
\multicolumn{1}{>{\columncolor[gray]{0.9}}c}{method} &
\multicolumn{1}{>{\columncolor[gray]{0.9}}c}{xc-functional} &
\multicolumn{2}{>{\columncolor[gray]{0.9}}c}{H$_2$O@O$_{\textit{br}}^{\textit{vac}}$}\\
\multicolumn{2}{>{\columncolor[gray]{0.9}}c}{} & 
\multicolumn{1}{>{\columncolor[gray]{0.9}}c}{A-TiO$_{2-\text{\sfrac{1}{4}}}$(101)} &
\multicolumn{1}{>{\columncolor[gray]{0.9}}c}{R-TiO$_{2-\text{\sfrac{1}{4}}}$(110)} \\[1mm]
\multirow{2}{*}{DFT} & PBE & 0.2 & 0.1$^a$\\
& HSE & 0.6 & 0.4$^a$ \\
\multirow{2}{*}{$G_0W_0$} & PBE & 0.7 & 0.6$^a$\\
& HSE & 1.0 & 0.9$^a$\\
\multicolumn{2}{c}{\multirow{2}{*}{STS}} & \multirow{2}{*}{$1.0\pm0.1^b$} & $0.7\pm0.1^b$\\
& & & $0.9^c$\\
\multicolumn{2}{c}{\multirow{2}{*}{PES}} & \multirow{2}{*}{$1.1^d$} & $0.9^d$\\
&  & & 0.8$^e$\\
\multicolumn{2}{c}{2PP} & & 0.9$^g$\\
\multicolumn{4}{p{0.95\columnwidth}}{\tiny$^a$Ref.~\citenum{MiganiH2O}. $^b$Ref.~\citenum{PhysRevLett.113.086402}. $^c$Ref.~\citenum{JinZhaoJCP2009STM}. $^d$Ref.~\citenum{PESXASRutileAnatasePRB2007}. $^e$Ref.~\citenum{DeSegovia}. $^g$Refs.~\citenum{PetekScienceH2O} and \citenum{PetekPRB2015}.}
\end{tabular}
\noindent{\color{StyleColor}{\rule{\columnwidth}{1.0pt}}}
\end{table}

Overall, \newRev{relative to $\Evac$, }the levels of A-TiO$_{2-\textrm{\sfrac{1}{4}}}$(101) are consistently about 0.6~eV lower in energy than those of R-TiO$_{2-\textrm{\sfrac{1}{4}}}$(110), for PBE DFT, HSE DFT, PBE $G_0W_0$, and HSE $G_0W_0$, as shown in Figure~\ref{fgr:ObrvacEvac}\textbf{(c-f)}.  However, the H$_2$O@O$_{\textit{br}}^{\textit{vac}}$ 1b$_2$ levels are at similar energies (within 0.2~eV) on A-TiO$_{2-\textrm{\sfrac{1}{4}}}$(101) and  R-TiO$_{2-\textrm{\sfrac{1}{4}}}$(110), for PBE DFT, HSE DFT, PBE $G_0W_0$, and HSE $G_0W_0$, as shown in Figure~\ref{fgr:ObrvacEvac}\textbf{(c-f)}.

\begin{figure}[!t]
\includegraphics[width=0.993\columnwidth]{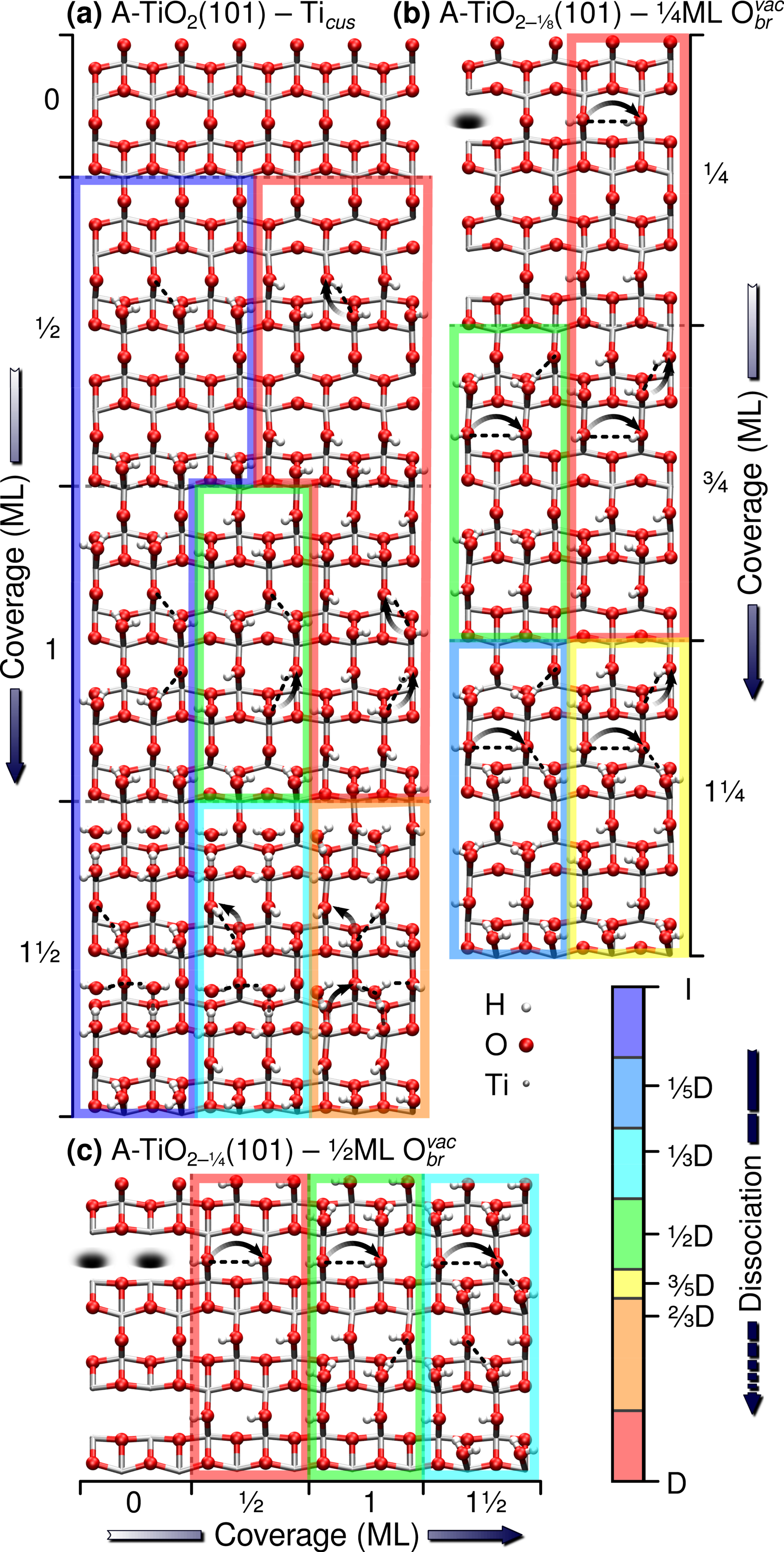}
\caption{Schematics of H$_2$O adsorbed intact (I) or dissociated (D) on 
\textbf{(a)} coordinately unsaturated Ti sites (Ti$_{\textit{cus}}$) of stoichiometric A-TiO$_2$(101) and \textbf{(b)} \sfrac{1}{4}ML or \textbf{(c)} \sfrac{1}{2}ML of bridging O vacancies (O$_{\textit{br}}^{\textit{vac}}$) on defective A-TiO$_{2-x}$(101), where $x$ = \sfrac{1}{8} and \sfrac{1}{4}, respectively. Coverage is half the number of H$_2$O formula units per (101) $1\times1$ unit area of the clean stoichiometric or defective surface. Dissociation is the fraction of H$_2$O molecules which are dissociated.  Charge transfer of about $-0.4e$ accompanying deprotonation is represented by arrows, while intermolecular (gray) and interfacial (black) hydrogen bonds are denoted by dotted lines.}\label{fgr:Schematics} 
\noindent{\color{StyleColor}{\rule{\columnwidth}{1pt}}}
\end{figure}

Focusing on the $\textit{IP}$ from PBE $G_0W_0$ \newRev{shown in Table~\ref{IPPBEG0W0:tbl}}, they are ordered: R-TiO$_2$(110) (7.29~eV) $\approx$ A-TiO$_2$(101) (7.15~eV) $>$ H$_2$O@Ti$_{\textit{cus}}$ on   A-TiO$_2$(101) (6.25~eV) $\approx$ H$_2$O@O$_{\textit{br}}^{\textit{vac}}$ on   A-TiO$_{2-\textrm{\sfrac{1}{4}}}$(101) (6.19~eV) $\approx$ H$_2$O@Ti$_{\textit{cus}}$ on   R-TiO$_2$(110) (6.03~eV) $>$ H$_2$O@O$_{\textit{br}}^{\textit{vac}}$ on   A-TiO$_{2-\textrm{\sfrac{1}{4}}}$(101) (5.37~eV). On the one hand, there are no significant differences in $\textit{IP}$ between bare and H$_2$O@Ti$_{\textit{cus}}$ covered A-TiO$_2$(101) and R-TiO$_2$(110).  On the other hand, for H$_2$O dissociatively adsorbed on O$_{\textit{br}}^{\textit{vac}}$ sites of defective A-TiO$_{2-\textrm{\sfrac{1}{4}}}$(101) and R-TiO$_{2-\textrm{\sfrac{1}{4}}}$(110), the $\textit{IP}$s are significantly different, despite having quite similar HO$_{\textit{br}}$ concentrations per unit area.  The origin of this difference might be related to differences in the structure's relative stability or surface dipole.\cite{JinH2O}

\begin{table}
\vspace{-9mm}
\noindent{\color{StyleColor}{\rule{\columnwidth}{1.0pt}}}
\vspace{4.5mm}
\caption{\newRev{\textrm{\bf{Ionization Potentials \textit{IP} in eV from PBE $\boldsymbol{G_0W_0}$ for A-TiO$_{\text{2}}$(101) and R-TiO$_{\text{2}}$(110)}}}
}\label{IPPBEG0W0:tbl}
\begin{tabular}{ccc}
\multicolumn{3}{>{\columncolor[gray]{0.9}}c}{ }\\[-3mm]
\multicolumn{1}{>{\columncolor[gray]{0.9}}c}{coverage} &
\multicolumn{1}{>{\columncolor[gray]{0.9}}c}{surface} & 
\multicolumn{1}{>{\columncolor[gray]{0.9}}c}{\textit{IP} (eV)}\\[1mm]
\multirow{2}{*}{clean} & A-TiO$_2$(101) & 7.15\\
 & R-TiO$_2$(110) & 7.29$^a$\\
\multirow{2}{*}{1ML H$_2$O@Ti$_{\textit{cus}}$} & A-TiO$_2$(101) & 6.25\\
 & R-TiO$_2$(110) & 6.03$^b$\\
\multirow{2}{*}{1ML H$_2$O@O$_{\textit{br}}^{\textit{vac}}$} & A-TiO$_{2-\sfrac{1}{4}}$(101) & 6.19\\
 & R-TiO$_{2-\sfrac{1}{4}}$(110) & 5.37$^b$\\\multicolumn{3}{p{0.95\columnwidth}}{\tiny$^a$Ref.~\citenum{MiganiLong}. $^b$Ref.~\citenum{MiganiH2O}.}
\end{tabular}
{\color{StyleColor}{\rule{\columnwidth}{1.0pt}}}
\end{table}

Similarly, $\epeak$ for dissociatively adsorbed H$_2$O@O$_{\textit{br}}^{\textit{vac}}$ on A-TiO$_{2-\textrm{\sfrac{1}{4}}}$(101) is about 0.4~eV below that on R-TiO$_{2-\textrm{\sfrac{1}{4}}}$(110).  Since $\epeak$ is thus closer to the standard hydrogen electrode (SHE) for  H$_2$O@O$_{\textit{br}}^{\textit{vac}}$ on R-TiO$_{2-\textrm{\sfrac{1}{4}}}$(110) than A-TiO$_{2-\textrm{\sfrac{1}{4}}}$(101), one would expect the former structure to require a smaller overpotential and be more active than the latter within an electrochemical cell \cite{NorskovSHEJPCB2004}.    However, \oldRev{photocatalytically,}\newRev{for photocatalysis,} the alignment of $\epeak$ relative to $\eVBM$ is the more relevant quantity.  As we shall see in the next section, the relative electrochemical and photocatalytic activities of these two structures are \oldRev{completely }reversed.

\subsection{Coverage and Dissociation Dependence of H$_{\text{2}}$O Spectra for Stoichiometric and Defective Surfaces}\label{Sect:Dependence}

To systematically investigate the similarities and differences between A-TiO$_2$(101) and R-TiO$_2$(110) surfaces, we consider a variety of coverages of intact and dissociated H$_2$O on stoichiometric A-TiO$_2$(101) \oldRev{(}\newRev{[}Figure~\ref{fgr:Schematics}\textbf{(a)}\oldRev{)}\newRev{]}  and defective A-TiO$_{2-x}$(101) \oldRev{(}\newRev{[}Figure~\ref{fgr:Schematics}\textbf{(b,c)}\newRev{]}, as done previously for the rutile surface \cite{MiganiH2O}.  These configurations are consistent with previous results for H$_2$O on A-TiO$_2$\cite{waterLatersAnatase,WaterlayerDefectAnatase,Anatase101WaterMD,Selloni1998219,SelloniWaterReview2010,TiloccaJCP2003,YadongPRLH2OSubsurface,DieboldSelloniNatMater2009,PhysRevLett.81.2954,DieboldJPCB2003}.  

\begin{table}[!t]
  \caption{\newRev{\textrm{\bf Adsorption Energies $\boldsymbol{E_{\textit{ads}}}$, Highest PDOS Peaks $\boldsymbol{\epeak}$ and Average PDOS HOMO Energies $\eHOMOPDOS$ in eV of H$_{\text{2}}$O on Ti$\boldsymbol{_{\textit{cus}}}$ of Stoichiometric A-TiO$_{\text{2}}$(101) and R-TiO$_{\text{2}}$(110) and O$\boldsymbol{_{\textit{br}}^{\textit{vac}}}$ of Defective A-TiO$\boldsymbol{_{2-x}}$(101) and R-TiO$\boldsymbol{_{2-x}}$(110) with \textit{x} = \sfrac{1}{8} or \sfrac{1}{4}.}}
}\label{Eads:tbl}
\begin{tabular}{c@{\ \ }c@{\ \ }cccc@{\ \ }cc@{\ \ \ \ \ }r@{.}l}
\multicolumn{10}{>{\columncolor[gray]{0.9}}c}{ }\\[-3mm]
\multicolumn{3}{>{\columncolor[gray]{0.9}}c}{coverage} &
\multicolumn{3}{>{\columncolor[gray]{0.9}}c}{A-TiO$_{2-x}$(101)} &
\multicolumn{4}{>{\columncolor[gray]{0.9}}c}{R-TiO$_{2-x}$(110)} \\[1mm]
\multicolumn{3}{>{\columncolor[gray]{0.9}}c}{ML\ \ \ \ \ \ \ \   $x$}&
\multicolumn{1}{>{\columncolor[gray]{0.9}}c}{$E_{\textit{ads}}$}&
\multicolumn{2}{>{\columncolor[gray]{0.9}}c}{$\eHOMOPDOS$  $\epeak$}&
\multicolumn{1}{>{\columncolor[gray]{0.9}}c}{$E_{\textit{ads}}$}&
\multicolumn{3}{>{\columncolor[gray]{0.9}}c}{$\eHOMOPDOS$  $\epeak$}\\[1mm]
\multirow{2}{*}{\sfrac{1}{2}} & I & 0 & -0.38 & -1.42& 
-0.7 
 & -0.33 & -1.44& 
-1&2 
\\
 & D & 0  & -0.18 & -0.49& 
-0.7 
&  -0.13 & -1.03& 
-0&8 
\\
\multirow{3}{*}{1} & I &  0 & -0.37 & -1.07& 
-0.6 
&-0.41 & -1.28& 
-1&1$^a$ 
\\
 & \sfrac{1}{2}D &  0 & -0.29 & -0.43& 
-0.5
& -0.23 & -0.79& 
-1&0$^a$
\\
 & D & 0  & -0.26 & -0.45& 
-0.5 
& -0.10 & -0.68& 
-0&9$^a$
\\
\multirow{3}{*}{1\sfrac{1}{2}} & I &  0 & -0.35 & -0.72& 
-0.7
& -0.34 & -0.99& 
-1&3 
\\
 & \sfrac{1}{3}D & 0  & -0.27 & -0.42& 
-0.7
& -0.17& -0.68& 
-1&1 
\\
 & \sfrac{2}{3}D &  0 & -0.24 & -0.39& 
-0.8
& -0.12 & -0.65& 
-0&9
\\[1mm]
\sfrac{1}{4} & D & \sfrac{1}{8} & -1.35 & -1.27& 
-1.0 
&-0.83 & -1.37& 
-1&1
 \\
 \multirow{2}{*}{\sfrac{3}{4}} & \sfrac{1}{3}D & \sfrac{1}{8} &  -0.69 & -1.04& 
-0.8 
&-0.44 & -1.15& 
-1&1
\\
 & D & \sfrac{1}{8} & -0.60 & -0.58& 
-0.7 
&-0.34 & -0.79& 
-0&8 
\\
 \multirow{2}{*}{1\sfrac{1}{4}} & \sfrac{1}{5}D & \sfrac{1}{8} &  -0.50 & -1.00& 
-0.6
&-0.47 & -1.15& 
-1&0 
\\
  & \sfrac{3}{5}D & \sfrac{1}{8} & -0.43 & -0.53& 
 -0.6 
& -0.40 & -0.75& 
 -1&0
\\[1mm]
 \sfrac{1}{2} & D & \sfrac{1}{4} & -1.35 & -0.92& 
-0.6 
&-1.32 & -1.18& 
-1&1$^a$ 
\\
 1 & \sfrac{1}{2}D & \sfrac{1}{4} & -0.68 & -0.90& 
-0.6
&-0.77 & -1.12& 
-1&1 
\\
 1\sfrac{1}{2} & \sfrac{1}{3}D & \sfrac{1}{4} & -0.55 & -0.99& 
 -0.6 
& -0.69 & -1.22& 
-1&0 
\\[1mm]
\multicolumn{10}{p{0.95\columnwidth}}{\tiny$^a$Ref.~\citenum{MiganiH2O}.}
\end{tabular}
{\color{StyleColor}{\rule{\columnwidth}{1.0pt}}}
\end{table}

\begin{figure*}
\includegraphics[width=\textwidth]{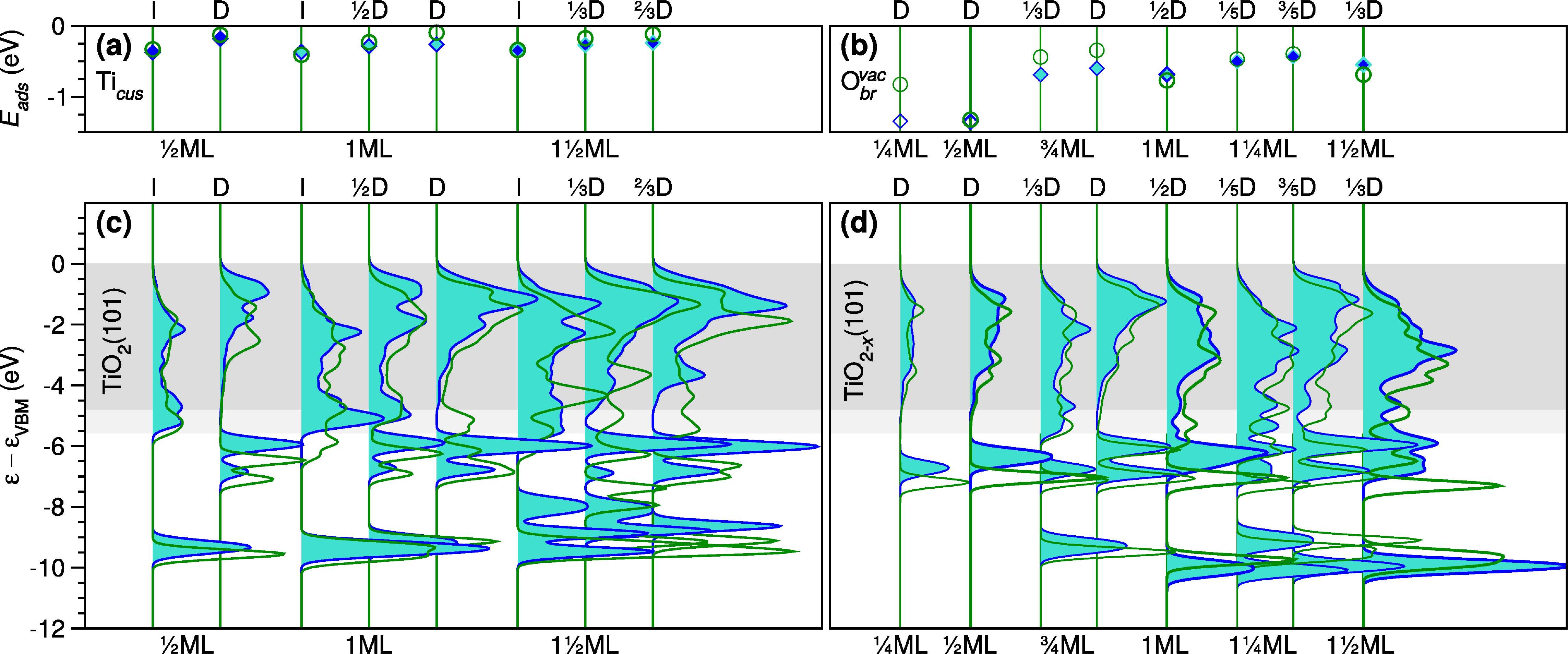}
\caption{
Structure and coverage dependence of \textbf{(a,b)} adsorption energy $E_{\textit{ads}}$ and \textbf{(c,d)} $G_0W_0$ PDOS for H$_{\text{2}}$O adsorbed intact (I) or dissociated (D) on  \textbf{(a,c)} coordinately unsaturated Ti sites (Ti$_{\textit{cus}}$) of stoichiometric A-TiO$_{\text{2}}$(101) (blue, Figure~\ref{fgr:Schematics}\textbf{(a)}) and R-TiO$_{\text{2}}$(110)\cite{MiganiH2O} (green) and \textbf{(b,d)} bridging O vacancies (O$_{\textit{br}}^{\textit{vac}}$) of defective A-TiO$_{2-x}$(101) (blue) and R-TiO$_{2-x}$(110)\cite{MiganiH2O} (green), with $x$ = \sfrac{1}{8} (thin lines, Figure~\ref{fgr:Schematics}\textbf{(b)}) or \sfrac{1}{4} (thick lines, Figure~\ref{fgr:Schematics}\textbf{(c)}). \textbf{(a,b)} RPBE $E_{\textit{ads}}$ on A-TiO$_{2-x}$(101) ($\meddiamond$) and R-TiO$_{2-x}$(110)\cite{MiganiH2O} ($\medcirc$) surfaces ($x$ = 0, \sfrac{1}{8}, \sfrac{1}{4}) for (white) low (\sfrac{1}{4} and \sfrac{1}{2}ML), (turquoise) medium (\sfrac{3}{4} and 1ML), and (blue) high (1\sfrac{1}{4} and 1\sfrac{1}{2}ML) coverage. \textbf{(c,d)} Energies are relative to the VBM ($\eVBM$).  The clean surface DOS of \textbf{(c)} A-TiO$_2$(101)/R-TiO$_2$(110) (dark/light gray regions) are shown for comparison. 
}\label{fgr:Spectra} 
\noindent{\color{StyleColor}{\rule{\textwidth}{1pt}}}
\end{figure*}

The adsorption energies shown in \newRev{Table~\ref{Eads:tbl} and }Figure~\ref{fgr:Spectra}\textbf{(a,b)} for H$_2$O on A-TiO$_2$, A-TiO$_{2-\textrm{\sfrac{1}{4}}}$, and A-TiO$_{2-\textrm{\sfrac{1}{8}}}$ (101) are generally similar to those on R-TiO$_2$, R-TiO$_{2-\textrm{\sfrac{1}{4}}}$, and R-TiO$_{2-\textrm{\sfrac{1}{8}}}$ (110)\cite{MiganiH2O}, respectively. On  both A-TiO$_2$(101) and R-TiO$_2$(110), intact H$_2$O adsorption is more stable than dissociative adsorption from \sfrac{1}{2} and 1\sfrac{1}{2} ML coverages.  The adsorption energies for H$_2$O@Ti$_{\textit{cus}}$ on A-TiO$_2$(101) follow the same trend as on R-TiO$_2$(110), but are somewhat stronger on A-TiO$_2$(101), with the greatest differences seen for dissociatively adsorbed H$_2$O.  Since the photocatalytically active species HO$_{\textit{br}}$@Ti$_{\textit{cus}}$ is more stable on A-TiO$_2$(101) than R-TiO$_2$(110), this also suggests that A-TiO$_2$(101) should be more photocatalytically active than R-TiO$_2$(110).  This is because one expects there to be more HO$_{\textit{br}}$@Ti$_{\textit{cus}}$ on A-TiO$_2$(101) than R-TiO$_2$(110).

On the defective surfaces, intact H$_2$O adsorption is also more stable than dissociative adsorption on A-TiO$_{2-\textrm{\sfrac{1}{4}}}$(101)/R-TiO$_{2-\textrm{\sfrac{1}{4}}}$(110) and A-TiO$_{2-\textrm{\sfrac{1}{8}}}$(101)/R-TiO$_{2-\textrm{\sfrac{1}{8}}}$(110).  Dissociative H$_2$O@O$_{\textit{br}}^{\textit{vac}}$ adsorption is generally stronger on A-TiO$_{2-\textrm{\sfrac{1}{4}}}$(101)/A-TiO$_{2-\textrm{\sfrac{1}{8}}}$(101) than R-TiO$_{2-\textrm{\sfrac{1}{4}}}$(110)/R-TiO$_{2-\textrm{\sfrac{1}{8}}}$(110) surfaces, except for 1\sfrac{1}{2}ML \sfrac{1}{3}D H$_2$O@O$_{\textit{br}}^{\textit{vac}}$.  

However, the adsorption energies shown in Figure~\ref{fgr:Spectra}\textbf{(b)} are strongly dependent on the stability of the defective A-TiO$_{2-\textrm{\sfrac{1}{4}}}$(101)/R-TiO$_{2-\textrm{\sfrac{1}{4}}}$(110) and A-TiO$_{2-\textrm{\sfrac{1}{8}}}$(101)/R-TiO$_{2-\textrm{\sfrac{1}{8}}}$(110) structures with surface O$_{\textit{br}}^{\textit{vac}}$.  Since surface O$_{\textit{br}}^{\textit{vac}}$ are less stable than subsurface O vacancies for A-TiO$_2$(101), the adsorption energies on A-TiO$_{2-x}$(101) provided in Figure~\ref{fgr:Spectra}\textbf{(b)} are somewhat overestimated.  

Figure~\ref{fgr:Spectra}\textbf{(c,d)} shows the \oldRev{QP }PBE $G_0W_0$ H$_2$O PDOS relative to $\eVBM$ as a function of coverage and dissociation for the structures shown in Figure~\ref{fgr:Schematics}.  Overall the PDOS on A-TiO$_2$(101) and R-TiO$_2$(110) are in surprisingly close agreement, both in shape and energy.  For \sfrac{1}{2}ML of H$_2$O, peaks related to the H$_2$O 1b$_2$, 3a$_1$ and 1b$_1$ levels (\emph{cf.} Figure~\ref{fgr:TicusEvac}\textbf{(b)}) are clearly evident. When the coverage is increased to more than 1ML, there are larger differences \oldRev{in }between the H$_2$O PDOS on A-TiO$_2$(101) and R-TiO$_2$(110). This \oldRev{is because of}\newRev{may be attributed to} the different intermolecular and interfacial interactions induced by the different hydrogen bonding networks. For 1\sfrac{1}{2} ML H$_2$O on A-TiO$_2$(101), the peak associated with the H$_2$O 1b$_2$ level, which is located at $-8$~eV, is more delocalized than rutile.  This is because there are more interfacial interaction\newRev{s} between H$_2$O and A-TiO$_2$ (101). For 1\sfrac{1}{4}ML H$_2$O on O$_{\textit{br}}^{\textit{vac}}$(\sfrac{1}{5}D), the water 1b$_2$ level splits into two peaks\oldRev{. This is because}\newRev{, as} H$_2$O and HO$_{\textit{br}}$ form two lines of hydrogen bonding networks.  We also find the bottom of the valance band for H$_2$O on A-TiO$_2$, A-TiO$_{2-\textrm{\sfrac{1}{4}}}$, and A-TiO$_{2-\textrm{\sfrac{1}{8}}}$ (101) is higher than that of R-TiO$_2$, R-TiO$_{2-\textrm{\sfrac{1}{4}}}$, and R-TiO$_{2-\textrm{\sfrac{1}{8}}}$ (110).  This is \oldRev{due}\newRev{attributable} to the higher VBM of the clean A-TiO$_2$(101) surface, as depicted by the gray regions in Figure~\ref{fgr:Spectra}. 

\begin{figure}[!t]
\includegraphics[width=\columnwidth]{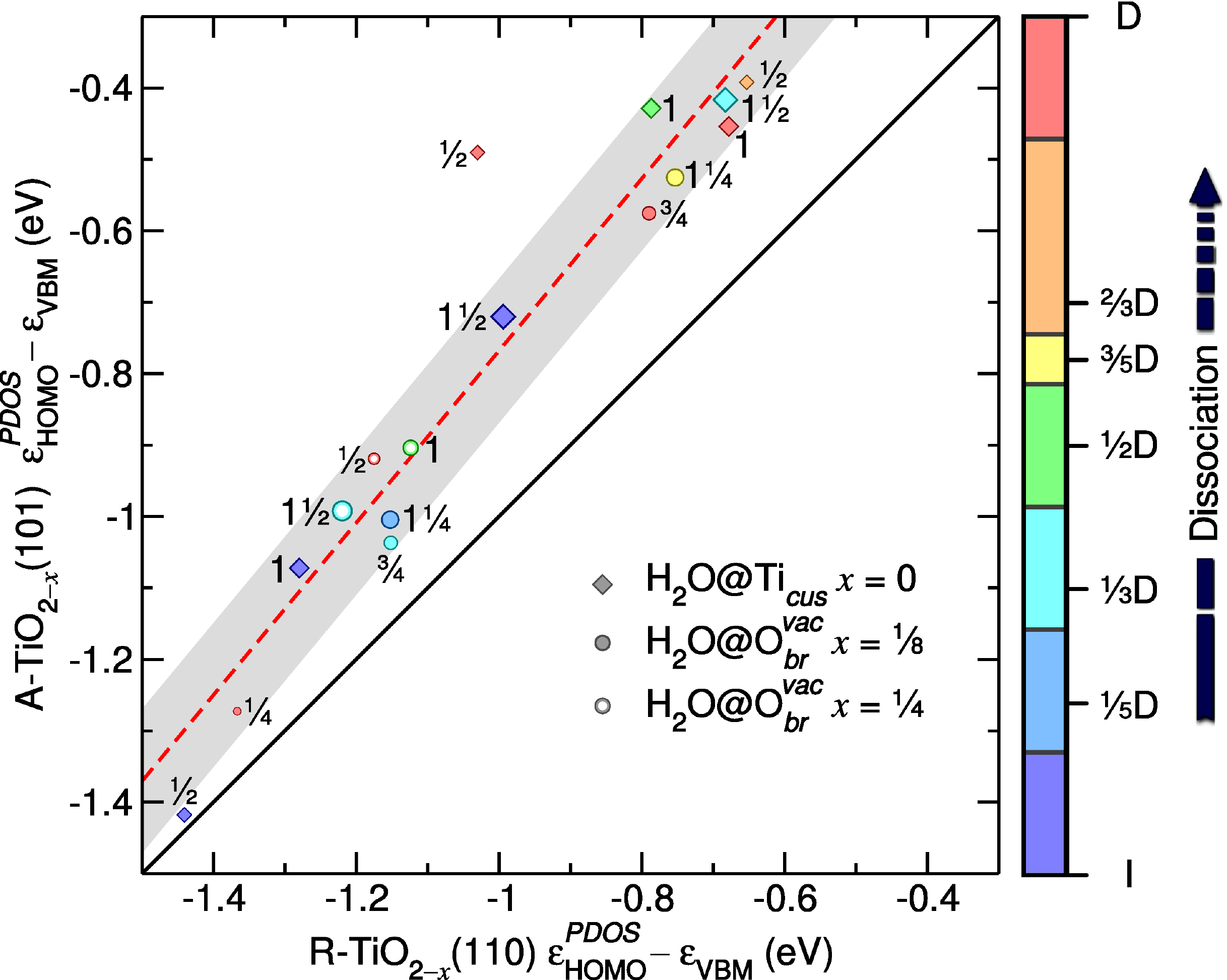}
\caption{\newRev{Average energy of \oldRev{QP }$G_0W_0$ PDOS HOMO $\eHOMOPDOS$ in eV of H$_2$O@Ti$_{\textit{cus}}$ on stoichiometric A-TiO$_{2}$(101) versus R-TiO$_{2}$(110) and of H$_2$O@O$_{\textit{br}}^{\textit{vac}}$ on defective  A-TiO$_{2-x}$(101) versus R-TiO$_{2-x}$(110) for $x$ = \sfrac{1}{8} or \sfrac{1}{4}. H$_{\text{2}}$O total coverage in ML and fraction intact (I) or dissociated (D) are provided. A linear fit (red dashed line) with a standard deviation of $\pm0.1$ eV (gray regions) is compared to the identity line (black solid line).}}
\label{fgr:EPDOSHOMO}
\noindent{\color{StyleColor}{\rule{\columnwidth}{1pt}}}
\end{figure}

For all spectra shown in Figure~\ref{fgr:Spectra}\textbf{(c,d)}, \oldRev{$\epeak$ }\newRev{$\eHOMOPDOS$} is higher on A-TiO$_2$, A-TiO$_{2-\textrm{\sfrac{1}{4}}}$, and A-TiO$_{2-\textrm{\sfrac{1}{8}}}$ (101) than R-TiO$_2$, R-TiO$_{2-\textrm{\sfrac{1}{4}}}$, and R-TiO$_{2-\textrm{\sfrac{1}{8}}}$ (110), respectively\newRev{, as shown in Table~\ref{Eads:tbl} and Figure~\ref{fgr:EPDOSHOMO}}.
  Further, these differences in \oldRev{$\epeak$}\newRev{$\eHOMOPDOS$} are larger for dissociated H$_2$O.  Since it is the HOMO of HO@Ti$_{\textit{cus}}$ which can trap a photogenerated hole, as discussed in section~\ref{Sect:Intact}, the larger differences in \oldRev{$\epeak$}\newRev{$\eHOMOPDOS$} shown in Figure\newRev{s}~\ref{fgr:Spectra}\textbf{(c)} \newRev{and \ref{fgr:EPDOSHOMO}} for dissociated H$_2$O suggest A-TiO$_2$(101) should generally be more photocatalytically active than R-TiO$_2$(110) from low coverage (\sfrac{1}{2}ML H$_2$O) to multi-layered H$_2$O (1\sfrac{1}{2}ML H$_2$O).

For \sfrac{1}{2}ML of dissociatively adsorbed H$_2$O@O$_{\textit{br}}^{\textit{vac}}$, $\epeak$ relative to the VBM for R-TiO$_{2-\sfrac{1}{4}}$(110) is below that for A-TiO$_{2-\sfrac{1}{4}}$(101).  This suggests HO$_{\textit{br}}$@O$_{\textit{br}}^{\textit{vac}}$ should be more photocatalytically active on A-TiO$_{2-\textrm{\sfrac{1}{4}}}$(101) compared to R-TiO$_{2-\textrm{\sfrac{1}{4}}}$(110).  However, as shown in the previous section, the reverse is true for their relative electrochemical activity, i.e., HO$_{\textit{br}}$@O$_{\textit{br}}^{\textit{vac}}$ on R-TiO$_{2-\textrm{\sfrac{1}{4}}}$(110) is expected to be more electrochemically active than A-TiO$_{2-\textrm{\sfrac{1}{4}}}$(101).  This demonstrates the importance of considering both the absolute level alignment relative to $\Evac$, and the level alignment relative to $\eVBM$. 

\section{CONCLUSIONS}\label{Sect:Conclusions}

\newRev{In heterogeneous catalysis, photocatalytic activity is controlled by the level alignment of the adsorbate and substrate levels. For this reason it is essential to obtain a quantitative description of the interfacial level alignment to determine and predict catalytic activity. This can only be obtained from many-body QP $GW$ calculations, which are necessary to correctly describe the anisotropic
screening of electron-electron interactions at the catalyst’s interface.  }

\newRev{Previously, we have shown that HSE $G_0W_0$ reliably describes the interfacial level alignment relative to the VBM for highly hybridized and  localized molecular levels of H$_2$O\cite{MiganiH2O} and CH$_3$OH\cite{MiganiLong}  on R-TiO$_2$(110). Here, we have shown that HSE $G_0W_0$ also provides a quantitative description of the occupied Ti$^{3+}$ $3d$ level's alignment relative to the Fermi level on both reduced anatase and rutile polymorphs.  These are the levels from which electrons are typically excited in 2PP experiments\cite{PetekPRB2015,Petek2PPH2O,PetekScienceH2O}. Since HSE DFT fails in both cases, these results clearly demonstrate the important role played by anisotropic screening of the electron-electron interaction in describing the alignment of these molecular and defect levels.
}

In this study we have performed an in-depth comparison of the QP $G\oldRev{_0}W\oldRev{_0}$  level alignment for H$_2$O--A-TiO$_2$(101) and H$_2$O--R-TiO$_2$(110) interfaces \oldRev{across a plethora of}\newRev{for a range of chemically significant} structures.  We have considered the limits of low and high H$_2$O coverage, intact to fully dissociated H$_2$O, and stoichiometric to O defective surface composition.  \newRev{Using the HOMO--VBM level alignment for these systems prior to irradiation $\eHOMOPDOS$, we have established the following trends in their relative photocatalytic activity for H$_2$O photooxidation.  (1) There is a strong linear correlation between $\eHOMOPDOS$ on A-TiO$_{2-x}$(101) and R-TiO$_{2-x}$(110).  (2) }We consistently find H$_2$O's $\eHOMOPDOS$  closer to $\eVBM$ for A-TiO$_2$ than R-TiO$_2$.  \newRev{(3) These differences in $\eHOMOPDOS$ are greater for dissociated H$_2$O, and increase as $\eHOMOPDOS$ approaches $\eVBM$.  (4) Overall, $\eHOMOPDOS$ approaches $\eVBM$ with H$_2$O dissociation.  Altogether, this}\oldRev{This} suggests \newRev{HO@Ti$_{\textit{cus}}$ is more photocatalytically active than intact H$_2$O@Ti$_{\textit{cus}}$ and} hole trapping is more favorable on A-TiO$_2$(101) than R-TiO$_2$(110)\oldRev{, and}\newRev{. This} may explain why the anatase polymorph is generally more photocatalytically active than rutile for H$_2$O photooxidation.  

\newRev{We have clearly demonstrated that the ground state interfacial level alignment is a key factor in understanding the photocatalytic activity of TiO$_2$. Moreover, in general, knowledge of an interface’s ground state electronic structure can be used to establish trends for predicting photocatalytic activity.}


\section*{\large$\blacksquare$\normalsize\ 
AUTHOR INFORMATION}
\subsubsection*{Corresponding Author}
\noindent E-mail: duncan.mowbray@gmail.com (D.J.M.)
\subsubsection*{Notes} 
\noindent The authors declare no competing financial interest.
\section*{\large$\blacksquare$\normalsize\ 
ACKNOWLEDGMENTS} 

We acknowledge financial support from the China Scholarship Council (CSC),  the European Projects DYNamo (ERC-2010-AdG-267374)\oldRev{ and}\newRev{,} CRONOS (280879-2 CRONOS CP-FP7)\newRev{, Cost Actions CM1204 (XLIC), and MP1306 (EuSpec)}; Spanish Grants (FIS2012-37549-C05-0\newRev{2}, FIS201\oldRev{0-21282-C02-01}\newRev{3-46159-C3-1-P}, PIB2010US-00652, RYC-2011-09582, JCI-2010-08156); Generalitat de Catalunya (2014SGR301, XRQTC); Grupos Consolidados UPV/EHU del Gobierno Vasco (IT-578-13); NSFC (21003113 and 21121003); MOST (2011CB921404); \newRev{the Chinese Academy of Sciences President's International Fellowship; }and NSF Grant CHE-1213189; and computational time from the Shanghai Supercomputing Center, BSC Red Espanola de Supercomputacion, and EMSL at PNNL by the DOE.
\bibliography{bibliography}


\end{document}